\documentclass{ptephy}

\usepackage{amsmath,amssymb,bm,float,Macro,mathrsfs,subfigure,tabularx,url}
\usepackage{threeparttable}

	\def\rmd{{\rm d}}
	\def\bk{\bm{k}}
	\def\hatk{\widehat{\bk}}
	\def\bl{\bm{\ell}}
	\def\bL{\bm{L}}
	\def\bn{\bm{\nabla}}
	\def\hatn{\hat{\bm n}}
	\def\ol{\overline}

	\def\grad{\phi}
	\def\curl{\varpi}
	\def\estx{\widehat{x}}
	\def\estg{\widehat{\grad}}
	
	\def\este{\widehat{\epsilon}}

	\def\tT{\widetilde{\Theta}}
	\def\tE{\widetilde{E}}
	\def\tB{\widetilde{B}}
	\def\tX{\widetilde{X}}
	\def\tY{\widetilde{Y}}

	\def\tXi{\widetilde{\Xi}}
	\def\hT{\widehat{\Theta}}

	\def\bT{\ol{\Theta}}

	\def\tC{\widetilde{C}}
	\def\hC{\widehat{C}}
	\def\lCTT{\tC^{\Theta\Theta}}
	\def\lCTE{\tC^{\Theta E}}
	\def\lCEE{\tC^{EE}}
	\def\lCBB{\tC^{BB}}
	\def\uCTT{C^{\Theta\Theta}}
	\def\uCTE{C^{\Theta E}}
	\def\uCEE{C^{EE}}
	\def\uCBB{C^{BB}}
	\def\hCTT{\hC^{\Theta\Theta}}

\def\hn{\widehat{n}}
\def\chis{\chi_{\rm s}}
\def\CAMB{{\tt CAMB} \cite{Lewis:1999bs}}

\begin{document}


\title{
	Cosmology from weak lensing of CMB 
} 

\author{
	\name{\fname{Toshiya} \surname{Namikawa}}{1} 
}

\address{
	\affil{1}{Yukawa Institute for Theoretical Physics, Kyoto University, Kyoto 606-8502, Japan}
	\email{namikawa@yukawa.kyoto-u.ac.jp}
}

\begin{abstract} 
The weak lensing effect on the cosmic microwave background (CMB) induces distortions in 
spatial pattern of CMB anisotropies, and statistical properties of CMB anisotropies 
become a weakly non-Gaussian field. 
We first summarize the weak lensing effect on the CMB (CMB lensing) 
in the presence of scalar, vector and tensor perturbations. 
Then we focus on the lensing effect on CMB statistics and methods to estimate 
deflection angles and their power spectrum. 
We end by summarizing recent observational progress and future prospect.
\end{abstract} 


\maketitle


\section{Introduction} 

The path of CMB photons emitted from the last scattering surface 
of CMB is deflected by
gravitational potential of the large-scale structure with typically a few arc-minute deflection.
This leads to the distortion in spatial pattern of observed CMB anisotropies. Among various
cosmological observations, a measurement of weak lensing signals in CMB maps is a direct
probe of intervening gravitational fields along a line of sight, and is considered as one of the
most powerful probes of fundamental issues in cosmology and physics in the near future.

Most of the pioneering work in CMB lensing focused on how the lensing effect modifies 
the two-point statistics of CMB temperature anisotropies 
(e.g., \cite{Blanchard:1987AA,Sasaki:1989,Tomita:1989,Fukushige:1994,Seljak:1995ve} and Refs therein). 
An accurate calculation of lensing effect on the angular power spectrum by
Ref.~\cite{Seljak:1995ve} showed that the acoustic scale imprinted in temperature is 
slightly smoothed and the small scale temperature fluctuations are enhanced by transferring 
large scale power to small scale. 
On the other hand, Refs.~\cite{Bernardeau:1996aa,Zaldarriaga:2000ud} showed 
that the lensing effect also modifies the statistics of CMB anisotropies and 
generates non-Gaussian signatures in the observed CMB anisotropies. 
A more interesting and important effect for future studies of CMB lensing is 
that the gravitational lensing generates B-mode polarization converted from E-mode
polarization \cite{Zaldarriaga:1998ar}.

Although the theoretical framework of CMB lensing has been established a decade ago, 
a significant observational progress has been made only recently. 
The lensing effect on CMB temperature and polarization anisotropies as well as
the gravitational lensing potential are now measured with both ground-based and satellite
experiments (e.g., \cite{Reichardt:2008ay,Das:2011ak,Hanson:2013daa} and see Sec. 
\ref{Sec:obs} for details), requiring studies of more practical
issues which are now rapidly developing. 
The measured lensing signals are already used for some specific issues in cosmology, e.g., 
dark energy \cite{Sherwin:2011gv,vanEngelen:2012va,Ade:2013tyw,Ade:2013zuv}, 
dark matter \cite{Wilkinson:2013kia}, 
cosmic strings \cite{Namikawa:2013wda}, and primordial non-Gaussianity \cite{Giannantonio:2013kqa}. 
Although statistical significance of the current detections of lensing signals is not so high 
compared to other cosmological probes, 
the lensing signals obtained from upcoming and next generation experiments will have enough potential 
to probe the following fundamental issues: 

\bi 
\item 
{\it Dark energy, Dark matter and Massive neutrinos:} 
The theoretical understanding of the nature of the dark energy is still limited, and 
the cosmological observations are the only way to reveal the dynamical properties of the dark energy. 
On the other hand, determination of the neutrino mass is one of the most important subjects 
in elementary particle physics, and is the key to understand the physics beyond the standard model 
of particle physics. The properties of the dark energy, specific models of dark matter, and mass of
neutrinos affect the evolution of gravitational potential, and thus the signals of weak lensing.
\vs{0.5}

\item 
{\it Gravitational waves, Cosmic strings and Magnetic fields:} 
A measurement of the curl mode of deflection angles is also interesting for cosmology 
(see Sec.~\ref{Sec:WL}). 
The curl-mode deflection angles are produced by the vector and tensor metric perturbations, 
but not by the scalar perturbations. 
That is, the non-vanishing curl-mode signal is a smoking gun of the non-scalar metric perturbations 
which can be sourced by gravitational waves \cite{Cooray:2005hm,Namikawa:2011cs} 
and cosmic strings \cite{Namikawa:2011cs,Yamauchi:2012bc,Yamauchi:2013fra} 
which may give clues about the mechanism of inflationary scenario at the early universe 
and implications for high-energy physics. 
The magnetic fields at cosmological scales would also be probed with the curl mode which 
will be explored in our future work. 
\ei

In addition, the study of CMB lensing has implications for detecting the signature of the
primordial gravitational waves, since the amplitude of CMB B-mode polarization generated
from primordial gravitational waves is smaller than that from lensing if the tensor-to-scalar
ratio is very small \cite{Knox:2002pe}. 
In order to enhance sensitivity to the primordial B-mode polarization, subtraction of 
the lensed B-mode would become important 
\cite{Verde:2005ff,Smith:2008an,Smith:2010gu,Abazajian:2013vfg}. 
Similarly, since the lensing induces the non-Gaussian signatures in CMB anisotropies and non-zero 
off-diagonal elements in the covariance matrix, the lensing effect would be a possible confusing
source in estimating the primordial non-Gaussianity or testing the statistical isotropy. 
For this reason, precise and accurate estimations of lensing effect on CMB maps are required.

As the measurements of lensing effect become more precise, the studies of CMB lensing should 
focus more on practical issues rather than on purely theoretical issues. 
For example, reconstruction of gravitational potential, which is mainly used for analysis of CMB 
lensing, is based on the assumption that the primordial CMB anisotropies are statistically isotropic. 
There are, however, several possible sources to generate mode couplings in the anisotropies 
such as the mask of Galactic emission and point sources \cite{Carvalho:2010rz,BenoitLevy:2013bc},  
inhomogeneous noise, \cite{Hanson:2009dr}, beam asymmetry \cite{Hanson:2010gu}, and so on. 
These contaminations potentially lead to a significant bias in the estimation of lensing potentials. 
For accurate cosmology with future observations, methods for mitigating all these biases are needed. 

This paper is organized as follows. 
In Sec.~\ref{Sec:WL}, we formulate the weak gravitational lensing in the presence 
of scalar, vector and tensor metric perturbations, and see how the lensing
signals depend on cosmological sources. 
In Sec.~\ref{Rev:CMBlensing}, we show how the lensing effect modifies the statistics of 
the observed CMB anisotropies. 
In Sec.~\ref{LensRec}, we discuss the method for estimating the deflection angles and 
their power spectrum. 
Sec.~\ref{Sec:obs} is devoted to summary of recent observational status and future prospect.

\section{Weak gravitational lensing from scalar, vector and tensor perturbations} 
\label{Sec:WL}

Consider a photon emitted from the last scattering surface of CMB, which passes through 
gravitational fields before reaching us. 
The geodesic of the photon is perturbed by the gravitational lensing, and the
photon is observed in a different direction from the original direction. 
The difference between observed and original directions is called the deflection angle 
which provides information on the anisotropies of projected gravitational fields 
integrated from the last scattering surface to the observer. 
In this section, we review how the deflection angle is related to 
the gravitational fields and how its power spectrum depends on properties of several cosmological
sources such as the dark energy, massive neutrinos and cosmic strings.

\subsection{Gradient and curl modes of deflection angle} 

The expression for the deflection angle with the metric perturbations is obtained by solving 
the photon geodesic in a perturbed universe. Let us consider the line element given by 
\al{ 
	\rmd s^2 = a^2(\eta) (\bar{g}_{\mu\nu}+\delta g_{\mu\nu})\rmd x^{\mu}\rmd x^{\nu}
	\,, \label{Eq:ds-perturbed} 
} 
where $a$ is the scale factor in a homogeneous and isotropic universe, $\bar{g}_{\mu\nu}$ is 
the background unperturbed metric and $\delta g_{\mu\nu}$ is the small metric perturbations. 
Here we assume that the unperturbed metric is described by the flat 
Friedman-Lema\^{i}tre-Robertson-Walker metric: 
\al{ 
	\bar{g}_{\mu\nu} \rmd x^{\mu}\rmd x^{\nu} 
		= - \rmd \eta^2 + \bar{\gamma}_{ij} \rmd x^i \rmd x^j 
		= - \rmd \eta^2 + \rmd \chi^2 + \chi^2\omega_{ab}\rmd \theta^a \rmd \theta^b
	\,, \label{Eq:ds-unperturbed} 
} 
with $\omega_{ab}\rmd\theta^a\rmd\theta^b=\rmd\theta^2+\sin^2\theta\rmd\varphi^2$ denoting the metric 
on the unit sphere. In the conformal Newton gauge, the metric perturbations are described as 
\footnote{
Note that Ref.~\cite{Yamauchi:2013fra} gives the derivation of deflection angles 
in terms of the gauge-invariant variables in linear perturbation theory \cite{Kodama:1984}.  
}
\al{ 
	\delta g_{00} &= - 2\Phi 
	\,, & 
	\delta g_{0i} &= - \sigma_{i} 
	\,, & 
	\delta g_{ij} &= 2\Psi\bar{\gamma}_{ij} + h_{ij} 
	\,, \label{ds-perturbed} 
} 
where the quantities, $\Phi$ and $\Psi$, are the scalar components, $\sigma_i$ is the divergence-free 
vector component ($\sigma_{i}{}^{|i}=0$), and $h_{ij}$ is the transverse-traceless tensor component 
($h_{ij}{}^{|i}=0$, and $h^i{}_{i}=0$). The vertical bar ($|$) denotes the covariant derivative 
with respect to the background three-dimensional metric, $\bar{\gamma}_{ij}$. 

To define the deflection angle, let us consider null geodesics in the background and perturbed 
spacetime, $\bar{x}^{\mu}$ and $x^{\mu}$. Since the photon path is not deflected in the background 
spacetime, the unperturbed path can be parametrized as $\bar{x}^{\mu}=(\eta_0-\chi,\chi\hn^i)$. 
Here the quantity $\eta_0$ denotes the conformal time today and $\hn^i$ is the unit vector describing 
the observed direction of photon. Assuming a static observer, we define the deflection angle by 
projecting the angular components of the deviation vector $x^{\mu}-\bar{x}^{\mu}$ on the sphere 
\cite{Yamauchi:2013fra}: 
\al{
	d^a \equiv \frac{[x^{i}(\chis)-\bar{x}^{i}(\chis)] e_{i}{}^{a}}{\chis} - \theta^a_O 
	\,, 
}
where the subscript $a$ means the angular components, $\theta$ and $\varphi$. The three-dimensional 
vectors, $e_{i}{}^a$, are the basis vectors orthogonal to $\hn^i$, the quantity, $\chis$, is the 
conformal distance between the observer and the last scattering surface of CMB, and $\theta^a_O$ is 
the angular coordinate at the observer. Since the deflection angle has two degrees of freedom, we 
decompose the deflection angle into two components by parity symmetry as 
\cite{Stebbins:1996wx,Hirata:2003ka,Cooray:2005hm,Namikawa:2011cs} 
\al{
	d^a = \grad^{:a} + \epsilon^a{}_b\curl^{:b} 
	\,, \label{Eq:LCMB:remap}
}
where $(:)$ is the covariant derivative on the unit sphere, and $\epsilon^a{}_b$ denotes the 
two-dimensional Levi-Civita symbol. Hereafter, we call the first and second terms in the right-hand 
side of Eq.~(\ref{Eq:LCMB:remap}) gradient and curl modes, respectively. 

The deviation vector is obtained from the geodesic equation in the perturbed spacetime, 
and the resultant expressions for the gradient and curl modes are given by \cite{Yamauchi:2013fra}
\al{
	&\bn^2\grad = d^a{}_{:a} 
		= \INT{}{\chi}{\chi}{0}{\chis} 
		\biggl\{\frac{\chis -\chi}{\chis}\bn^2[2\psi+\mC{A}] - \mC{B}^a{}_{:a}\biggr\}
	\,, \label{WL:geo-eq:grad} 
	\\ 
	&\bn^2\curl = d^a{}_{:b}\,\epsilon^b{}_a
		= - \INT{}{\chi}{\chi}{0}{\chis}\mC{B}^a{}_{:b}\,\epsilon^b{}_a
	\,. \label{WL:geo-eq:curl}
}
Here $\bn^2$ is the Laplacian operator on the unit sphere, $\psi=(\Phi-\Psi)/2$, and we define the 
quantities generated by non-scalar perturbations: 
\al{
	\mC{A} = \sigma_i \widehat{n}^i + h_{ij}\widehat{n}^i\widehat{n}^j 
	\,, \ \ \
	\mC{B}_a = \sigma_i e^i_a - 2 h_{ij}e^i_a\widehat{n}^j 
	\,. \label{Eq:lens-metric}
}
In Eqs.~\eqref{WL:geo-eq:grad} and \eqref{WL:geo-eq:curl}, the integral at the right-hand-side is 
evaluated along the unperturbed light path, usually referred to as the Born approximation 
(see e.g. Ref.~\cite{Cooray:2002mj} for the correction terms). 
The radial displacement (or the time delay) is also discussed in Ref.~\cite{Hu:2001yq}, 
but is a negligible effect on statistical observables. 


Eq.~\eqref{WL:geo-eq:curl} shows that the curl mode of the deflection angle vanishes 
if we consider the scalar perturbations alone, 
but it is produced by the vector and/or tensor components. 
Also, beyond the liner perturbation, the curl mode is generated, e.g., 
by the second order of the density perturbations \cite{Hirata:2003ka,Sarkar:2008ii}. 

It is also worth noting about the relation between the deflection angle and the elements of the Jacobi 
matrix which is defined as the mapping between a source and an image plane. 
The relation between the deflection angle and the Jacobi matrix is obtained by solving 
the geodesic deviation equation. Denoting the symmetric-traceless part of the Jacobi matrix 
divided by $\chi\rom{s}$ as $\gamma_{ab}$ (shear components), 
the relation becomes \cite{Schmidt:2012nw,Yamauchi:2013fra} 
\al{
	\gamma_{ab} = d_{\ave{a:b}} + \frac{1}{2}[h_{\ave{ab}}]_0^{\chi_s} 
	\,, 
}
where, for any quantity, $X_{\ave{ab}}=(X_{ab}+X_{ba}-X^c{}_c\omega_{ab})/2$. 
The last term arises in the presence of tensor perturbations as a difference of coordinate system 
perturbed by the metric at observer and source position \cite{Dodelson:2003bv}.

\subsection{Angular power spectrum of gradient and curl modes} 

Once we measure the deflection angle, one of the useful quantities in cosmology is the
angular power spectrum of fluctuations, rather than the fluctuations themselves. 
Here we turn to discuss the angular power spectrum of the gradient and curl modes, 
based on Eqs.~\eqref{WL:geo-eq:grad} and \eqref{WL:geo-eq:curl}. 
To see how the observable depends on properties of the cosmological sources, we also
show the relation between the angular power spectrum and the power spectrum of metric
perturbations.

\subsubsection{Scalar perturbations alone} 

Let us first consider the case in the presence of scalar perturbations alone, 
since it helps our understanding of derivation in the presence of the non-scalar perturbations.

The fluctuations of gravitational potential are decomposed into Fourier modes 
with the scalar-mode function $Q^{(0)}(\bm{x},\bk)=\E^{-\iu\bk\cdot\bm{x}}$ as 
\al{
	\psi(\bm{x},\eta) 
		&= \Int{3}{\bk}{(2\pi)^3} \psi_{\bk}(\eta) Q^{(0)}(\bm{x},\bk)
	\label{Eq:grad:scalar} 
	\,. 
}
Substituting the above equation into Eq.~\eqref{WL:geo-eq:grad}, we obtain 
\al{
	\bn^2 \grad 
		= 2\INT{}{\chi}{_}{0}{\chis}\frac{\chis -\chi}{\chis\chi} 
		\Int{3}{\bk}{(2\pi)^3} \psi_{\bk}(\eta) \bn^2 Q^{(0)}(\chi\hatn,\bk) 
	\,. \label{WL:geo-eq:grad:1} 
}

The angular power spectrum is defined as 
\al{
	\delta_{\ell\ell'}\delta_{mm'}C_{\ell}^{\grad\grad} = \ave{\grad_{\ell m}\grad_{\ell' m'}^*}
	\,, \label{Eq:def-Clpp}
}
where the quantity $\grad_{\ell m}$ is the spherical harmonic coefficients 
of the gradient mode and is defined with the spin-$0$ spherical harmonics $Y_{\ell m}(\hatn)$ as 
\al{
	\grad(\hatn) = \sum_{\ell,m}\grad_{\ell m}Y^*_{\ell m}(\hatn) 
	\,. 
}
Substituting the above equation into Eq.~\eqref{WL:geo-eq:grad:1}, and using the orthogonality of the 
spherical harmonics, we obtain 
\footnote{
Note that we ignore $\ell=0$ mode since this mode produces the mean value of the gradient mode 
and is not observable using a measurement of deflection angles. 
}
\al{
	\grad_{\ell m} = \frac{2}{\ell(\ell+1)}\INT{}{\chi}{_}{0}{\chis}\frac{\chis -\chi}{\chis\chi}
		\Int{3}{\bk}{(2\pi)^3} \psi_{\bk}(\eta) \Int{2}{\hatn}{_} Y_{\ell m}(\hatn)
		\bn^2 Q^{(0)}(\chi\hatn,\bk)  
	\,, 
	\label{Eq:grad_ellm}
}
where we use $\bn^2Y_{\ell,m}=\ell(\ell+1)Y_{\ell,m}$. To simplify the above equation, we use 
\al{
	\Int{2}{\hatn}{_}Y_{\ell m}(\hatn)\bn^2 Q^{(0)}(\chi\hatn,\bk) 
		= 4\pi (-\iu)^{\ell}j_{\ell}(k\chi) \ell(\ell+1) Y^*_{\ell m}(\hatk)
	\label{Eq:grad:Q0} 
	\,, 
}
where $j_{\ell}(x)$ is the spherical Bessel function. 
Substituting the above equation into Eq.~\eqref{Eq:grad_ellm}, we obtain 
\al{
	\grad_{\ell m} = 2\INT{}{\chi}{_}{0}{\chis}\frac{\chis -\chi}{\chis\chi}
		\Int{3}{\bk}{2\pi^2} (-\iu)^{\ell}j_{\ell}(k\chi)
		\psi_{\bk}(\eta_0-\chi) Y^*_{\ell m}(\hatk) 
	\,, 
	\label{Eq:grad_Yellm}
}
With the dimensionless power spectrum defined as 
\al{
	\ave{\psi_{\bk}(\eta)\psi^*_{\bk'}(\eta')} 
		= (2\pi)^3\delta(\bk-\bk')\frac{2\pi^2}{k^3}\Delta_{\psi}(k,\eta,\eta') 
	\,, 
}
the angular power spectrum of the gradient mode defined in Eq.~\eqref{Eq:def-Clpp} is given by 
\al{
	C_{\ell}^{\grad\grad} 
		&= 16\pi \Int{}{k}{k} 
		\INT{}{\chi}{_}{0}{\chis}\INT{}{\chi'}{_}{0}{\chis}
		\frac{\chis -\chi}{\chis\chi}\frac{\chis -\chi'}{\chis\chi'}
	j_{\ell}(k\chi)j_{\ell}(k\chi') \Delta_{\psi}(k,\eta_0-\chi,\eta_0-\chi') 
	\,. \label{Eq:Clpp}
}

\subsubsection{General case} 

Next we consider the case in the presence of all types of the metric perturbations. 
Similar to Eq.~\eqref{Eq:grad:scalar}, the vector and tensor metric perturbations are also decomposed 
into Fourier modes with the mode functions of vector $Q^{(\pm 1)}_i(\bm{x},\bm{k})$ and tensor 
$Q^{(\pm 2)}_{ij}(\bm{x},\bm{k})$, respectively \cite{Hu:1997hp}: 
\al{
	\sigma_i(\bm{x},\eta) &= \Int{3}{\bk}{(2\pi)^3}\sum_{s=\pm 1}
		\sigma_{\bk}^{(s)}(\eta) Q^{(s)}_i(\bm{x},\bm{k}) 
	\,, \label{Eq:vec-decomp} \\ 
	h_{ij}(\bm{x},\eta) &= \Int{3}{\bk}{(2\pi)^3}\sum_{s=\pm 2}
		h_{\bk}^{(s)}(\eta) Q^{(s)}_{ij}(\bm{x},\bk) 
	\,, \label{Eq:tens-decomp}
}
where the explicit forms of these mode functions are given by \cite{Hu:1997hp}
\al{
	Q^{(\pm 1)}_i(\bm{x},\bk) 
		&=\frac{\pm\iu}{\sqrt{2}} e_{\pm,i}(\hatk)\E^{-\iu\bk\cdot\bm{x}} 
	\,, \\ 
	Q^{(\pm 2)}_{ij}(\bm{x},\bk) 
		&=\frac{-1}{\sqrt{2}}e_{\pm,i}(\hatk)e_{\pm,j}(\hatk)\E^{-\iu\bk\cdot\bm{x}} 
	\,. 
}
Here the polarization vector 
$\bm{e}_{\pm}(\hatk)=\bm{e}_{\theta}(\hatk)\pm\iu\bm{e}_{\varphi}(\hatk)$ is 
perpendicular to the wave vector $\bk$.  
Similar to the case of the scalar perturbations alone, we first substitute Eqs.~\eqref{Eq:vec-decomp} 
and \eqref{Eq:tens-decomp} into Eq.~\eqref{WL:geo-eq:grad} and \eqref{WL:geo-eq:curl}. 
From Eq.~\eqref{Eq:lens-metric}, we then compute the similar form of Eq.~\eqref{Eq:grad:Q0} but for, 
e.g., $\bn^2\widehat{n}^iQ^{(\pm 1)}_i$ instead of $\bn^2 Q^{(0)}$. 
More generally, what we must compute is a quantity $\mC{J}^{(G)}_{\ell}(k\chi)$ defined as 
\al{
	\Int{2}{\hatn}{_}Y_{\ell m}(\hatn) G(\chi\hatn,\bk)
		= \mC{J}^{(G)}_{\ell}(k\chi) Y^*_{\ell m}(\hatk)
	\label{Eq:G} 
	\,, 
}
where $G$ is for example $\bn^2\widehat{n}^iQ^{(\pm 1)}_i$. 
Ref.~\cite{Hu:1997hp} obtained the functional form of $\mC{J}^{(G)}$ 
(see also Ref.~\cite{Dai:2012bc,Yamauchi:2013fra}) 
and applied to the calculation of the CMB angular power spectrum. 
As shown in Ref.~\cite{Yamauchi:2013fra}, Eq.~\eqref{Eq:G} also simplifies the computation of 
the angular power spectra for the gradient and curl modes. 
To relate the angular power spectrum of the gradient and curl modes to the dimensionless power 
spectrum of the metric perturbations, we assume that the statistical properties of the vector and 
tensor modes are given by 
\al{
	&\ave{[\sigma^{(s)}_{\bk}(\eta )]^*\sigma_{\bk'}^{(s')}(\eta' )}
		= \delta_{ss'}\frac{1}{2}(2\pi)^3\delta(\bk-\bk') 
		\frac{2\pi^2}{k^3}\Delta^{(s)}(k;\eta ,\eta' ) 
	\qquad (s,s'=\pm 1) 
	\,, \\ 
	&\ave{[h_{\bk}^{(s)}(\eta )]^*h_{\bk'}^{(s')}(\eta' )}
		= \delta_{ss'}\frac{1}{8}(2\pi)^3\delta(\bk-\bk') 
		\frac{2\pi^2}{k^3}\Delta^{(s)}(k;\eta ,\eta' ) 
	\qquad (s,s'=\pm 2) 
	\,. 
}
The angular power spectrum of the curl mode $C_{\ell}^{\curl\curl}$ and the cross power spectrum 
between the gradient and curl modes $C_\ell^{\grad\curl}$ are also defined in the same form of 
Eq.~\eqref{Eq:def-Clpp}. The resultant angular power spectra are decomposed into the contributions 
from the scalar, vector and tensor perturbations as \cite{Yamauchi:2013fra}
\al{
	C_\ell^{xx} 
		&= 4\pi\INT{}{k}{k}{0}{\infty} k^2 \INT{}{\chi}{_}{0}{\chis}\INT{}{\chi'}{_}{0}{\chis} 
		\notag \\ 
	&\qquad \times\sum_{s=-2}^2 S^{(s)}_{x,\ell}(k\chis,k\chi)S^{(s)}_{x,\ell}(k\chis,k\chi') 
		\Delta^{(s)}\left( k;\eta_0 -\chi ,\eta_0 -\chi'\right)
	\,,\label{Eq:WL:Cl:GCpower} 
}
and $C_\ell^{\grad\curl}=0$, where $x=\grad$ or $\varpi$. The expressions of the transfer function, 
$S^{(s)}_{x,\ell}(\lambda,\lambda')$, are summarized as follows : 
\bi 
\item scalar perturbations  
\al{
	&S^{(0)}_{\grad,\ell}(\lambda,\lambda') 
		= 2\frac{\lambda-\lambda'}{\lambda\lambda'} j_{\ell}(\lambda') 
	\,,\label{Eq:WL:Cl:S-transfer-grad} \\ 
	&S^{(0)}_{\curl,\ell}(\lambda,\lambda') = 0 
	\,.\label{Eq:WL:Cl:S-transfer-curl} 
}
\item vector perturbations 
\al{
	&S^{(\pm 1)}_{\grad,\ell}(\lambda,\lambda') 
		= \sqrt{\frac{(\ell+1)!}{2(\ell-1)!}}
		\left[\frac{\lambda-\lambda'}{\lambda\lambda'} \frac{j_{\ell}(\lambda')}{\lambda'} 
			- \frac{1}{\ell(\ell+1)}\frac{1}{(\lambda')^2}\D{[\lambda'j_{\ell}(\lambda')]}{\lambda'}
		\right]
	\,,\label{Eq:WL:Cl:V-transfer-grad}\\ 
	&S^{(\pm 1)}_{\curl,\ell}(\lambda,\lambda') 
		= \pm \sqrt{\frac{1}{2\ell(\ell+1)}}j_{\ell}(\lambda')
		\,,\label{Eq:WL:Cl:V-transfer-curl} 
}
\item tensor perturbations 
\al{
	&S^{(\pm 2)}_{\grad,\ell}(\lambda,\lambda') 
		= \sqrt{\frac{(\ell+2)!}{32(\ell-2)!}}\left[\frac{\lambda-\lambda'}{\lambda\lambda'}
			\frac{j_{\ell}(\lambda')}{\lambda'} 
			- \frac{2}{\ell(\ell+1)}\frac{1}{(\lambda')^3}
			\D{[\lambda'j_{\ell}(\lambda')]}{\lambda'} 
		\right]
			+ \frac{\delta_{\ell,2}}{10\sqrt{3}}\delta(\lambda') 
	\,, \label{Eq:WL:Cl:T-transfer-grad} 
	\\ 
	&S^{(\pm 2)}_{\curl,\ell}(\lambda,\lambda') 
		= \pm \sqrt{\frac{(\ell+2)!}{(\ell-2)!}}
			\frac{1}{2\ell(\ell+1)} \frac{j_{\ell}(\lambda')}{\lambda'} 
	\,. \label{Eq:WL:Cl:T-transfer-curl} 
}
\ei 

\subsubsection{Angular power spectrum of gradient and curl modes} 

\begin{figure}
\bc
\includegraphics[width=90mm,clip]{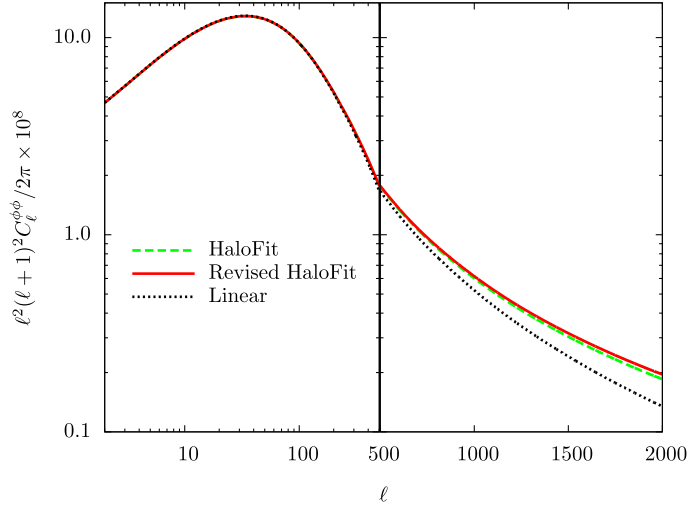}
\caption{
The angular power spectrum of the gradient mode generated by matter density fluctuations with the 
linear matter power spectrum (black dotted), and with fitting formula of the non-linear matter power 
spectrum given in Ref.~\cite{Smith:2002dz} (green dashed) or \cite{Takahashi:2012em} (red solid). 
}
\label{Fig:Clgg-Clcc}
\ec
\end{figure}

\begin{figure}[t]
\bc
\includegraphics[width=75mm,clip]{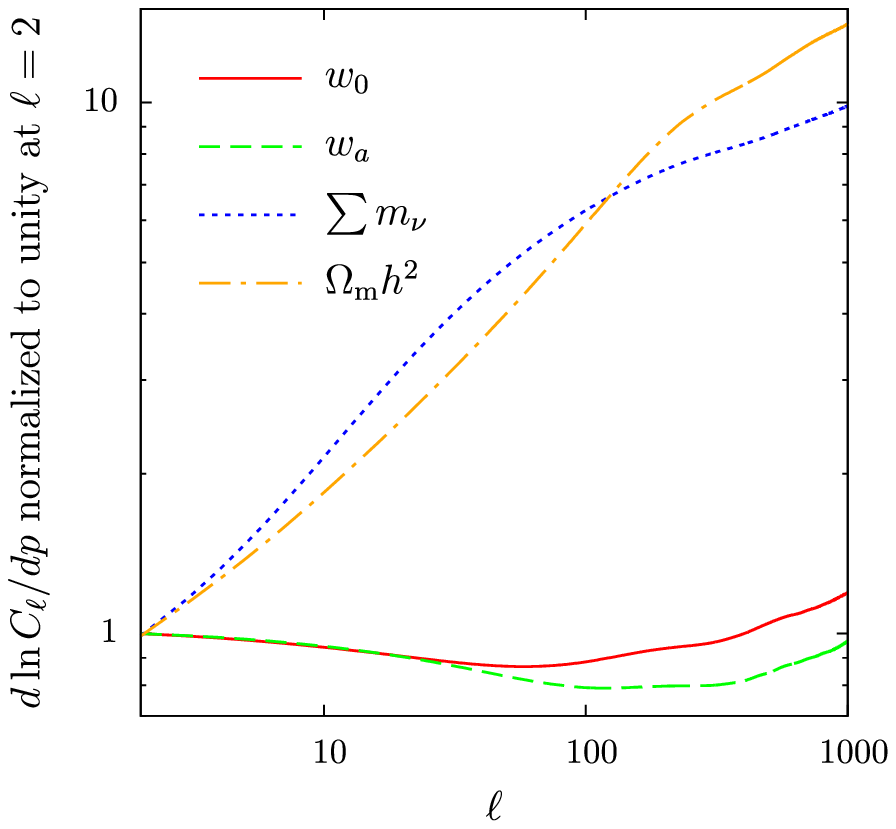}
\includegraphics[width=75mm,clip]{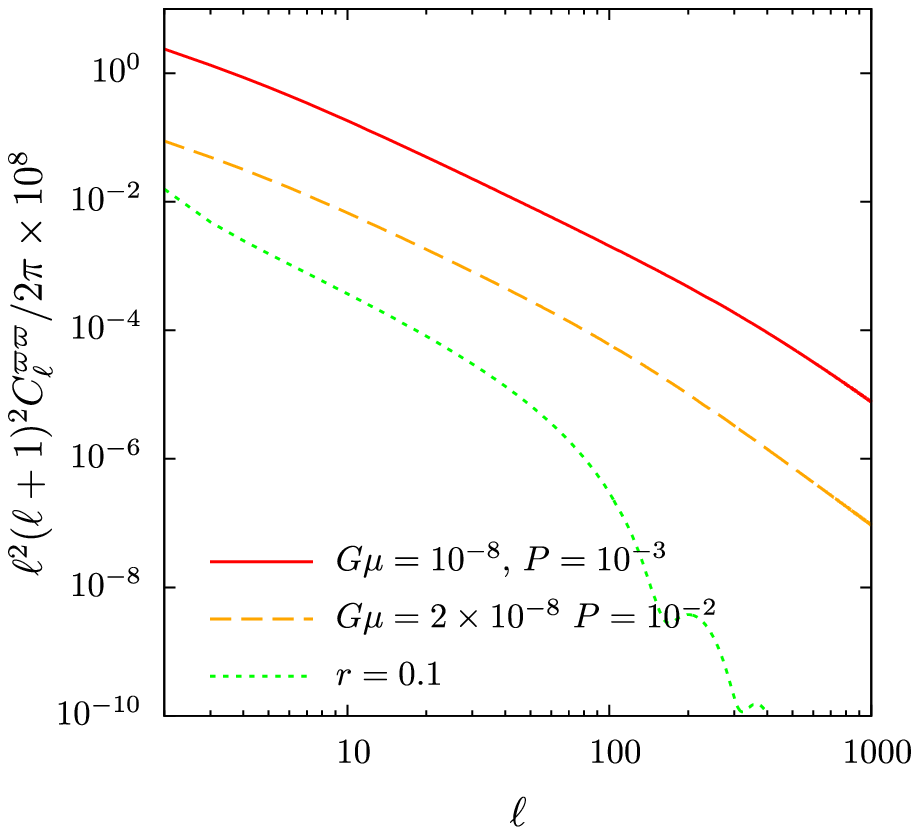}
\caption{
{\it Left} : Logarithmic derivatives of the gradient-mode power spectrum 
${\rm d}\ln C_{\ell}^{\grad\grad}/{\rm d}p$ with respect to the dark-energy equation-of-state 
parameters, $w_0$ (red solid), $w_a$ (green dashed), total mass of neutrinos, 
$\sum m_{\nu}$ (blue dotted) and $\Omega_{\rm m}h^2$ (orange long-dashed). 
The derivatives are normalized with the value at $\ell=2$. 
Note that the sign of the derivative with respect to $\Omega_{\rm m}h^2$ 
is positive, while the others have negative sign. 
{\it Right} : The angular power spectrum of the curl mode generated by 
the primordial gravitational waves with the tensor-to-scalar ratio $r=0.1$ (green dotted) and 
a specific model of the cosmic-string network (red solid/orange dashed). 
}
\label{fig:deriv}
\ec 
\end{figure}

Fig.~\ref{Fig:Clgg-Clcc} shows the angular power spectrum of the gradient mode 
generated by the matter density fluctuations. Three lines show the case with different fitting 
formulas of the matter power spectrum, i.e., the halofit model \cite{Smith:2002dz} and 
its revised formula 
\cite{Takahashi:2012em} in calculating angular power spectrum. For comparison, we also show the case 
with the linear power spectrum. Note that the lensing power spectrum is computed with \CAMB. 
The linear approximation to the matter power spectrum would be accurate at the scales where the 
signal becomes large ($\ell\sim 10-100$). The non-linear growth of matter density perturbations 
enhances the amplitude with $20$ - $30\%$ at $\ell\sim2000$ compared to linear theory. 
The sensitivity of $C_{\ell}^{\grad\grad}$ to the models of the non-linear evolution would be 
not so significant even at these scales, because the lensing power spectrum computed with 
the halofit model of Ref.~\cite{Smith:2002dz} is only a few percent smaller than the revised formula. 


In the left panel of Fig.~\ref{fig:deriv}, to see how the angular power spectrum depends on 
cosmological sources, we show the logarithmic derivatives of the angular power spectrum, 
$C_{\ell}^{\grad\grad}$ with respect to $w_0$ and $w_a$, a parameterization of the dark-energy 
equation-of-state as $w=w_0+(1-a)w_a$, and the total mass of neutrinos $\sum m_{\nu}$. 
For comparison, we also show dependence on the matter density $\Omega_{\rm m}h^2$. 
Note that the derivatives are normalized with a value at $\ell=2$. 
The derivatives with respect to the neutrino mass depend on $\ell$, 
since the presence of the massive neutrinos suppresses 
the matter density fluctuations at smaller scales than their free-streaming scale 
after they become non-relativistic particles \cite{1980:Bond}. 
On the other hand, the derivatives with respect to $w_0$ and $w_a$ are almost scale-independent
because the density fluctuations are affected by the properties of the dark energy through 
the evolution of the scale factor in the linear perturbation regime. 
These behaviors imply that the power spectrum of the gradient mode can distinguish 
the effect of the neutrino mass from that of the dark energy through 
the scale dependence \cite{Kaplinghat:2003bh}. 
We note however that there exist some parameters that exhibit a similar 
scale-dependence of the total neutrino mass, which can be a source of parameter degeneracy 
(see e.g., \cite{Namikawa:2010re}). 
As shown in Fig.~\ref{fig:deriv}, the logarithmic derivative with respect to 
the matter density gives a similar trend to that of the neutrino mass. 
This is because the matter density changes not only the amplitude of matter density fluctuations 
but also shift the peak of the matter power spectrum which is determined by 
the radiation-matter equality. 
Within the CMB data set, the degeneracy between the total mass of neutrinos and matter density 
remains and other external dataset would be required to break this degeneracy. 

On the other hand, in the right panel of Fig.~\ref{fig:deriv}, 
we show examples of the curl-mode angular power spectrum generated by 
the primordial gravitational waves with the tensor-to-scalar ratio $r=0.1$, 
and a specific model of cosmic string networks \cite{Yamauchi:2011cu} parametrized by 
the tension $G\mu$ and reconnection probability $P$. 
The angular power spectrum decreases at smaller scales since the perturbations are suppressed 
at sub-horizon scale. 
That is, a measurement of the curl-mode power spectrum on large scale is important 
to probe the primordial-gravitational waves and cosmic-string networks.

\section{Lensing effect on CMB anisotropies} \label{Rev:CMBlensing}

Lensing effect on CMB anisotropies modifies the statistical properties of the observed CMB 
anisotropies. The non-Gaussian behavior in the lensed anisotropies is particularly important 
for measuring the angular power spectrum of the gradient and curl modes as discussed in the next 
section. In this section, we briefly summarize the weak lensing effect on the angular power spectrum 
of CMB temperature and polarization, and non-Gaussian statistics such as the bispectrum and 
trispectrum, in the presence of both the gradient and curl modes 
(see also Ref.~\cite{Smith:2011we} for a review on the non-Gaussian statistics by lensing). 

\subsection{Lensed CMB angular power spectrum} 

The lensed temperature anisotropies $\tT(\hatn)$ are expressed as a remapping of unlensed temperature 
anisotropies $\Theta(\hatn)$ by a deflection angle (e.g., \cite{Blanchard:1987AA}): 
\al{ 
	\tT(\hatn) = \Theta (\hatn+\bm{d}(\hatn)) 
	\,. 
} 
The lensing effect on the CMB polarization is also described as a remapping of the Stokes parameters 
$Q\pm \iu U$ by the deflection angle. 

To analyze the asymptotic property of the lensed angular power spectrum, it is convenient to express 
the angular power spectrum in the flat-sky approximation. 
If we consider a small patch on a unit sphere and ignore the sky curvature, 
the CMB anisotropies are approximately given on a two-dimensional plane. 
In this limit, the lensed temperature anisotropies are expanded in terms of the plane wave: 
\al{
	\tT (\hatn) 
		&= \Int{2}{\bl}{(2\pi)^2}\tT_{\bl}\E^{\iu\bl\cdot\hatn} 
	\,. \label{Eq:LCMB:flat-T} 
}
On the other hand, since the lensed polarization anisotropies $\widetilde{Q}\pm\iu \widetilde{U}$ 
are the spin $\pm 2$ quantity, we define the rotationally invariant quantities, 
$\tE_{\bl}$ and $\tB_{\bl}$, usually referred to as E and B modes: 
\al{
	[\widetilde{Q}\pm \iu \widetilde{U}] (\hatn) 
		&= - \Int{2}{\bl}{(2\pi)^2}(\tE_{\bl}\pm\iu \tB_{\bl})
			\E^{\pm\iu 2\varphi_{\bl}} \E^{\iu\bl\cdot\hatn} 
	\,. \label{Eq:LCMB:flat-P} 
}
Here $\varphi_{\bl}$ is the azimuthal angle measured from the $x$-axis of the two dimensional plane. 
We can also expand the unlensed CMB anisotropies and define $\Theta_{\bl}$, $E_{\bl}$ and $B_{\bl}$ 
in the same way. 
The lensed (unlensed) angular power spectrum $\tC_{\ell}$ 
is then defined with the Fourier multipoles as 
\al{
	\ave{\tX_{\bl}\tY_{\bl'}} = (2\pi)^2 \delta(\bl-\bl') \tC_{\ell}^{XY} 
	\qquad (X,Y=\Theta,E,B)
	\,. 
}

A method to obtain the angular power spectrum is to expand the lensed anisotropies in terms of 
the deflection angle up to second order of the deflection angle \cite{Hu:2000ee}: 
\al{ 
	{}_s\tXi(\hatn) = {}_s\Xi (\hatn) + d^a {}_s\Xi_{:a}(\hatn) 
		+ \frac{d^ad^b}{2}{}_s\Xi_{:ab}(\hatn) 
	\,, \label{Eq:LCMB:remap-expand} 
} 
where $s=0$ or $\pm 2$ and we define ${}_{0}\Xi=\Theta$ and ${}_{\pm 2}\Xi=Q\pm\iu U$. 
The lensed CMB anisotropies in Fourier space become \cite{Hu:2000ee,Cooray:2005hm} 
\al{
	{}_{s}\tXi_{\bl} &= {}_{s}\Xi_{\bl} 
		- \sum_{x=\grad,\curl} \Int{2}{\bL}{(2\pi)^2} [(\bl-\bL)\odot_x\bL] x_{\bl-\bL}{}_{s}\Xi_{\bL} 
		\E^{\iu s\varphi_{\bL,\bl}} 
	\notag \\ 
		&+ \frac{1}{2}\sum_{x,y=\grad,\curl}\Int{2}{\bL}{(2\pi)^2}\Int{2}{\bL'}{(2\pi)^2}
			[\bL'\odot_x\bL][(\bl-\bL-\bL')\odot_y\bL'] x_{\bL'}y_{\bl-\bL-\bL'} 
			{}_{s}\Xi_{\bL}\E^{\iu s\varphi_{\bL,\bl}} 
	\,, \label{Eq:lensed-CMB}
}
where $\varphi_{\bL,\bl}=\varphi_{\bL}-\varphi_{\bl}$ and, for arbitrary two-dimensional vectors, 
$\bm{a}$ and $\bm{b}$, we define the products, $\odot_{\grad}$ and $\odot_{\curl}$, as 
\al{ 
	\bm{a}\odot_{\grad}\bm{b} \equiv a_{\theta}b_{\theta}+a_{\varphi}b_{\varphi} 
	\,, \qquad 
	\bm{a}\odot_{\curl}\bm{b} \equiv a_{\varphi}b_{\theta}-a_{\theta}b_{\varphi} 
	\,.
} 
Using Eq.~\eqref{Eq:lensed-CMB} and denoting the unlensed CMB angular power spectrum as 
$C_{\ell}$, the lensed angular power spectrum for temperature and polarization 
in the flat-sky approximation becomes \cite{Hu:2000ee,Cooray:2005hm} 
\al{
	\lCTT_{\ell} &= b_{\ell}\uCTT_{\ell} 
		+ \sum_{x=\grad,\curl}\Int{2}{\bL}{(2\pi)^2}[\bL\odot_x(\bl-\bL)]^2 C_{|\bl-\bL|}^{xx}\uCTT_L 
	\,, \label{Eq:lClTT} \\
	\lCTE_{\ell} 
		&= b_{\ell}\uCTE_{\ell} 
			+ \sum_{x=\grad,\curl}\Int{2}{\bL}{(2\pi)^2}[\bL\odot_x(\bl-\bL)]^2C_{|\bl-\bL|}^{xx} 
			\uCTE_L\cos 2\varphi_{\bL,\bl} 
	\,, \label{Eq:lClTE} \\ 
	\lCEE_{\ell} 
		&= b_{\ell}\uCEE_{\ell} + \sum_{x=\grad,\curl}
			\Int{2}{\bL}{(2\pi)^2} [\bL\odot_x(\bl-\bL)]^2 C_{|\bl-\bL|}^{xx} 
		\notag \\ 
		&\qquad\qquad\qquad\qquad \times \frac{1}{2}
			[(\uCEE_L+\uCBB_L)+(\uCEE_L-\uCBB_L)\cos 4\varphi_{\bL,\bl}] 
	\,, \label{Eq:lClEE} \\
	\lCBB_{\ell} 
		&= b_{\ell}\uCBB_{\ell} + \sum_{x=\grad,\curl}
		\Int{2}{\bL}{(2\pi)^2}[\bL\odot_x(\bl-\bL)]^2C_{|\bl-\bL|}^{xx}
		\notag \\ 
		&\qquad\qquad\qquad\qquad \times \frac{1}{2}
			[(\uCEE_L+\uCBB_L)-(\uCEE_L-\uCBB_L)\cos 4\varphi_{\bL,\bl}] 
	\,, \label{Eq:lClBB}
}
where we assume that correlation between the gradient and curl modes vanishes, and define 
\al{
	b_{\ell} = 1-\ell^2\sum_{x=\grad,\curl}\Int{}{\ln L}{4\pi}L^4 C_L^{xx} 
		= 1-\frac{\ell^2}{2}\ave{|\bm{d}|^2} 
	\,. 
} 
The lensing effect on the polarization has an interesting feature \cite{Zaldarriaga:1998ar}; 
even in the absence of the primary B-mode polarization, $C_{\ell}^{BB}=0$, 
spatial pattern of the lensed polarization anisotropies could have odd-parity mode. 
This is because a curl-free pattern is modified at each position and the resultant pattern 
is no longer a pure E-mode pattern. 
Note that a few arcminute deflection, $\ave{|\bm{d}|^2}\sim \mC{O}(10^{-7})$, \cite{Lewis:2006fu} 
leads to $\ell^2\ave{|\bm{d}|^2}/2\sim \mC{O}(1)$ if $\ell\sim 2500$. 
At these scales, the above expression is no longer valid because 
Eqs.~\eqref{Eq:lClTT}-\eqref{Eq:lClBB} ignore the higher order terms $\mC{O}(C_{\ell}^{xx})$, 
and a more accurate approach of Refs.~\cite{Seljak:1995ve,Zaldarriaga:1998ar} is required 
in which the angular power spectrum is computed using the correlation function, and 
the higher-order terms are included non-perturbatively with an exponential function 
\cite{Challinor:2005jy}. 

\begin{figure}
\bc
\includegraphics[width=78mm,clip]{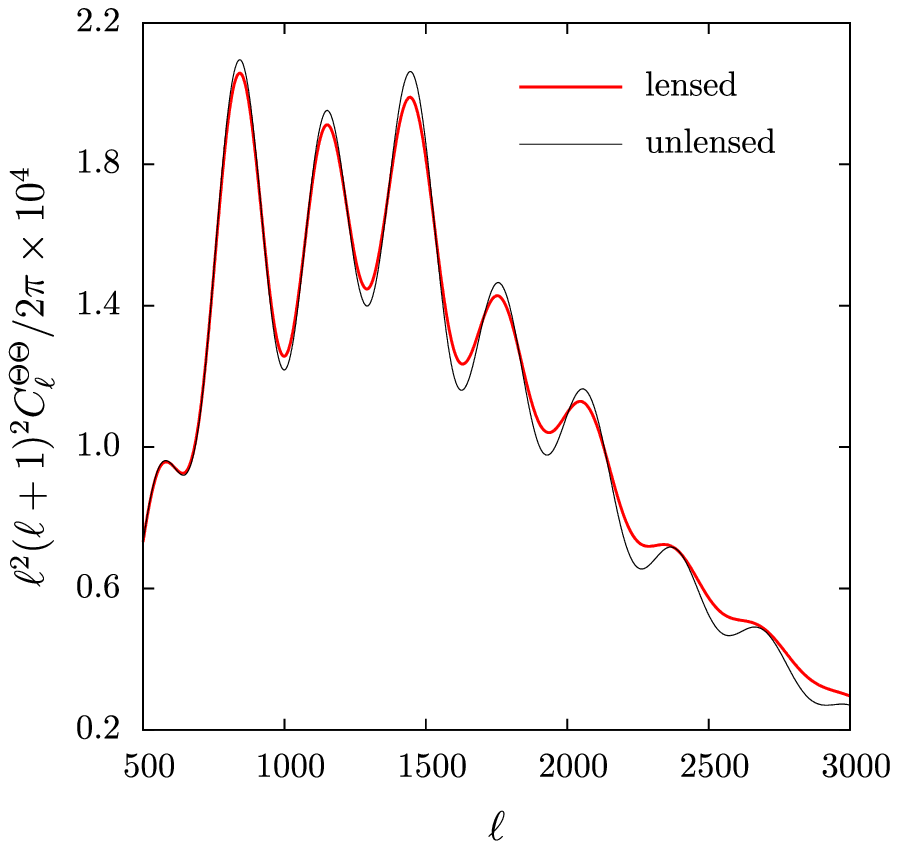}
\includegraphics[width=72mm,clip]{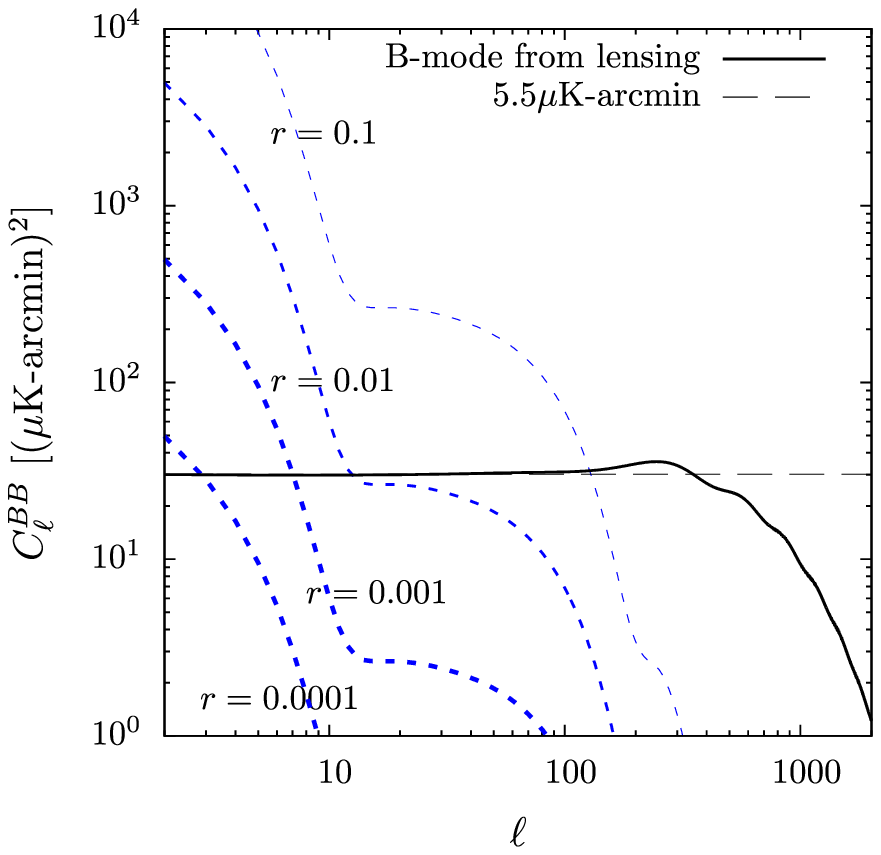}
\caption{
Summary of important signatures in the angular power spectra of the lensed anisotropies 
compared to those of the unlensed anisotropies. 
Note that contributions from the curl mode $C_{\ell}^{\varpi\varpi}$ is ignored. 
{\it Left} : 
Comparison of the angular power spectra between the lensed (red) and unlensed (black) 
temperature anisotropies. 
For illustrative purpose, we multiply $\ell^2(\ell+1)^2\times 10^4/2\pi$ to 
the angular power spectrum. 
{\it Right} : 
The angular power spectrum of the $B$-mode polarization from the primordial gravitational waves 
(blue dotted), compared with that from the lensing (black solid). 
For the primordial gravitational waves, the tensor-to-scalar ratio $r$ is varied from 
$0.1$ to $0.0001$. 
We also plot the white noise spectrum of $5.5\,\mu$K-arcmin (black dashed). 
}
\label{chap3-LensCl}
\ec
\end{figure}

In Fig.~\ref{chap3-LensCl}, we plot the lensed angular power spectrum of the temperature (left) and 
B-mode polarization (right), where the angular power spectrum is computed with \CAMB. 
The acoustic peaks in the temperature power spectrum are smeared by the lensing effect, 
and this can be understood as follows. 
The acoustic peaks are determined by typical sizes of hot and cold temperature spots in the sky. 
Lensing changes the size distribution of these spots (some get bigger and some get smaller), 
smearing the acoustic peaks. 
At small scales, the temperature power spectrum is dominated by lensing due to the transfer of 
large-scale power to small scales, where the primary temperature fluctuations are damped. 
The lensing also affects the E-mode power spectrum in the similar way of the temperature case. 

The B-mode power spectrum generated by the lensing effect is, on the other hand, 
a scale-independent spectrum on large scales ($\ell\,\lsim 100$) which roughly corresponds to 
$5$ - $6\mu$K-arcmin white noise. 
This would be considered as follows. 
The gradient of the primary E-mode has significant correlation on small scales, 
but does not correlate very much on scales larger than 
a few degree (or $\ell\sim 100$ \cite{Smith:2010gu}). 
%
The lensing only remaps these small scale fluctuations by typically a few arc-minute, 
and the resultant B-mode beyond degree scales 
is roughly an uncorrelated random field. 
This leads to the white noise spectrum on large scales. 

As shown in Fig.~\ref{chap3-LensCl}, 
the primary B-mode power spectrum generated by the primordial gravitational waves 
is smaller than the lensing B-mode at recombination bump ($\ell\sim 10$-$100$) 
if the tensor-to-scalar ratio is $r\,\lsim 0.01$, and at reionization bump ($\ell<10$) 
if $r\,\lsim 0.0001$ \cite{Knox:2002pe}. 
Therefore, if $r$ is small, detection of the primary B-mode requires subtraction of the 
lensing B-mode, so called delensing (see e.g., \cite{Seljak:2003pn,Teng:2011xc,Smith:2010gu}). 

\subsection{Non-Gaussian signatures of lensed CMB anisotropies} 

\subsubsection{Bispectrum} 

Assuming that the primary temperature anisotropies are a random Gaussian field, the three-point 
correlation, $\ave{\Theta_{\bl_1}\Theta_{\bl_2}\Theta_{\bl_3}}$, vanishes. 
As described in Eq.~\eqref{Eq:lensed-CMB}, the lensing, however, induces mode-couplings in 
the CMB anisotropies. 
The ISW-lensing correlation then generates the three-point correlation of lensed CMB anisotropies as 
\cite{Goldberg:1999xm} 
\al{
	\ave{\tT_{\bl_1}\tT_{\bl_2}\tT_{\bl_3}} 
		= (2\pi)^2\delta(\bl_1+\bl_2+\bl_3)
		[C_{\ell_1}^{\Theta \grad}\uCTT_{\ell_2}\bl_1\odot_{\grad}\bl_2 + ({\rm perms})] 
		+ \mC{O}(|d^a|^2) 
	\,. 
}
The ISW-lensing correlation generated by density perturbations on large scale, 
$C_{\ell}^{\Theta\grad}$, can be used to probe the late time evolution of the large-scale structure 
\cite{Seljak:1998nu,Goldberg:1999xm,Zaldarriaga:2000ud}. 
The cosmic strings at late time of the universe also produce the temperature bispectrum 
through $C_{\ell}^{\Theta\grad}$ \cite{Yamauchi:2013pna}, but contributions from the curl mode 
vanish since the cross correlation of temperature and the curl mode is an odd-parity quantity. 
The curl mode generated by the cosmic strings would be, however, a source of 
the polarization bispectrum through, e.g., $C_{\ell}^{B\curl}$. 
Note that the lensing bispectrum is a confusing source in estimating the primordial non-Gaussian 
signatures, especially for the squeezed type (e.g., $\ell_1\sim\ell_2\gg \ell_3$), 
and the bias for the local type becomes $f_{\rm NL}\sim \mC{O}(1)$ with ongoing and upcoming 
experiments \cite{Serra:2008wc,Cooray:2008xz,Hanson:2009kg}. 
The polarization bispectrum generated by lensing is also discussed in 
Ref.~\cite{Lewis:2011fk}.

\subsubsection{Trispectrum} 

The trispectrum of lensed CMB anisotropies has been explored 
\cite{Bernardeau:1996aa,Zaldarriaga:2000ud,Hu:2001fa} 
and can be used to estimate the power spectrum of the gradient and curl modes 
as discussed in the next section. 

The four-point correlation of CMB anisotropies is in general decomposed into two terms as 
\al{
	\ave{\tT_{\bl_1}\tT_{\bl_2}\tT_{\bl_3}\tT_{\bl_4}}
		= \ave{\tT_{\bl_1}\tT_{\bl_2}\tT_{\bl_3}\tT_{\bl_4}}_{\rm d} 
		+ \ave{\tT_{\bl_1}\tT_{\bl_2}\tT_{\bl_3}\tT_{\bl_4}}_{\rm c} 
	\,, \label{Eq:4-point-decomp}
}
where the first term is the disconnected part expressed in terms of two point correlations as 
\al{
	\ave{\tT_{\bl_1}\tT_{\bl_2}\tT_{\bl_3}\tT_{\bl_4}}_{\rm d}
		= \ave{\tT_{\bl_1}\tT_{\bl_2}}\ave{\tT_{\bl_3}\tT_{\bl_4}}
		+ (\bl_2\leftrightarrow \bl_3) 
		+ (\bl_2\leftrightarrow \bl_4) 
	\,, \label{Eq:4-point:d}
}
while the second term denotes the connected part which is not expressed in terms of 
the two point-correlation of lensed anisotropies alone 
and reflects the non-Gaussian behavior of the fluctuations. 
The four-point correlation of the lensed temperature anisotropies has a contribution from 
the connected part as \cite{Kesden:2003cc} 
\al{
	\ave{\tT_{\bl_1}\tT_{\bl_2}\tT_{\bl_3}\tT_{\bl_4}}\rom{c} 
	&\simeq (2\pi)^2\delta(\bl_1+\bl_2+\bl_3+\bl_4) 
	\notag \\ 
	&\qquad 
		\times \sum_{x=\grad,\curl}\left\{C_{|\bl_1+\bl_2|}^{xx}
			f^{x}_{\bl_1+\bl_2,\bl_1}f^{x}_{\bl_3+\bl_4,\bl_3}
			+ (\bl_2\leftrightarrow \bl_3) + (\bl_2\leftrightarrow \bl_4) 
		\right\}
	\,, \label{Eq:lens-trispectrum}
}
where we define a weight function as 
\al{ 
	f^{x}_{\bL,\bL'} = (\bL\odot_x\bL')\uCTT_{L'} + (\bL'\leftrightarrow \bL-\bL')
	\,. \label{Eq:weight} 
} 
In deriving Eq.~\eqref{Eq:lens-trispectrum}, the lensed temperature anisotropies are expanded only 
up to first order of the gradient and curl modes. 
The trispectrum of polarization generated by the lensing effect 
is also obtained analogously and the expression is given in Ref.~\cite{Okamoto:2002ik}.

\subsubsection{Other statistics} 

There are also several papers discussing how the non-Gaussian signatures of the lensing effect 
change the statistical properties of a random Gaussian field, 
such as topological statistics \cite{Schmalzing:2000,Takada:2001b}, and 
the two-point correlation of hot spots \cite{Takada:2000,Takada:2001a}. 
Non-zero lensing trispectrum also modifies the covariance of the lensed CMB angular power spectrum 
\cite{Smith:2005ue,Smith:2006nk,Li:2006pu,BenoitLevy:2012va}.

\section{Lensing Reconstruction} 
\label{LensRec}

Estimators for the lensing deflection fields in quadratic form of observed CMB anisotropies 
have been derived by several authors. 
Refs.~\cite{Zaldarriaga:1998te,Seljak:1998aq} developed a method for extracting lensing fields 
from temperature anisotropies with real space quantities. 
The method was subsequently extended to the case with polarization \cite{Guzik:2000ju}. 
The quadratic estimator mostly used in the recent analysis was developed in Fourier space 
by Refs.~\cite{Hu:2001,Hu:2001kj} and \cite{Okamoto:2003zw} in flat and full sky, respectively, 
and was also extended to include the curl mode by Ref.~\cite{Cooray:2005hm} in flat sky 
and by Ref.~\cite{Namikawa:2011cs} in full sky. 
On the other hand, the estimator is also derived in the context of the maximum likelihood 
\cite{Hirata:2002jy,Hirata:2003ka}. 
These estimators all utilize the fact that a fixed gradient/curl mode introduces statistical 
anisotropy into the observed CMB, in the form of a correlation between the CMB anisotropies and its 
gradient. 
With a large number of observed CMB modes, this correlation may be used to form estimates of the 
gradient and curl modes. 
The lensing power spectrum, which is required for cosmological analysis 
is then estimated from the gradient/curl mode estimators.

In this section, to see how to estimate the lensing power spectrum, $C_{\ell}^{\grad\grad}$ 
and $C_{\ell}^{\curl\curl}$, 
we first review the method for estimating the gradient and curl modes, usually referred to as lensing 
reconstruction, and their use of measuring the angular power spectrum 
(see also \cite{Hanson:2009kr} for a detailed review on the lensing reconstruction). 
Since the lensing fields would be measured more precisely in the near future, 
we also discuss the method for estimating lensing fields and their power spectrum 
with better accuracy. 

\subsection{Estimating CMB lensing potentials} 
\label{LensRec:Potential}

\subsubsection{Quadratic estimator for gradient and curl modes} 

For simplicity, let us first consider an estimator with a CMB temperature map alone 
in the absence of curl modes. In the following, observed temperature anisotropies and their angular 
power spectrum are denoted as $\hT_{\bl}$ and $\hCTT_{\ell}$, respectively. 
We assume that the observed anisotropies are given by $\hT_{\bl}=\tT_{\bl}+n_{\bl}$ where $n_{\bl}$ 
is the isotropic noise. 

The observed direction of the lensed CMB at each position is shifted from the original direction 
according to the deflection angle, and the distance 
between two positions are modified at each position in a different way. 
For a fixed realization of the deflection angle, the correlation function of the primary CMB 
anisotropies depends not only on the distance between two positions but also on 
the position in the sky. In Fourier space, the lensing-induced anisotropy leads to the 
correlations between two different Fourier modes of the observed CMB anisotropies. 
\al{
	\ave{\hT_{\bL}\hT_{\bl-\bL}}\rom{CMB,n} 
		= \grad_{\bl} f^{\grad}_{\bl,\bL} 
	\,, 
	\quad (\bl\not=\bm{0}) 
	\label{Eq:off-diag}
}
where $f^{\grad}_{\bl,\bL}$ is given in Eq.~\eqref{Eq:weight}, and 
we denote the ensemble average over primary CMB anisotropies and noise by $\ave{\cdots}\rom{CMB,n}$, 
to distinguish it from the usual meaning of the ensemble average, $\ave{\cdots}$. 

Based on Eq.~\eqref{Eq:off-diag}, the quantity, $\estg_{\bl,\bL}$, defined as  
\al{
	f^{\grad}_{\bl,\bL} \estg_{\bl,\bL} = \hT_{\bL}\hT_{\bl-\bL} 
	\,. 
}
is an estimator which satisfies the unbiased condition : 
$\ave{\estg_{\bl,\bL}}\rom{CMB,n}=\grad_{\bl}$ where $\bL$ is chosen so that 
$f_{\bl,\bL}^{\grad}\not=0$. 
.
A more optimal and unbiased estimator can be obtained as a sum of $\estg_{\bl,\bL}$ 
in terms of $\bL$, and the resultant estimator is \cite{Hu:2001}
\al{
	\estg_{\bl} 
		= A^{\grad}_{\ell}\Int{2}{\bL}{(2\pi)^2} g^{\grad}_{\bl,\bL} \hT_{\bL}\hT_{\bl-\bL} 
	\,. \label{Eq:QE-g}
}
Here the normalization, $A^{\grad}_{\ell}$, 
and the weight function, $g^{\grad}_{\bl,\bL}$ are given by 
\al{
	A^{\grad}_{\ell} = \frac{1}{[g^{\grad},f^{\grad}]_{\bl}} 
	\,; \quad 
	g^{\grad}_{\bl,\bL} 
		= \frac{f^{\grad}_{\bl,\bL}}{2\hCTT_{L}\hCTT_{|\bl-\bL|}} 
	\,, 
}
where, for convenience, the inner product is defined as 
\al{
	[g^x,f^y]_{\bl} \equiv \Int{2}{\bL}{(2\pi)^2} g^{x}_{\bl,\bL}f^{y}_{\bl,\bL} 
	\,. 
}

In the presence of the curl mode, Eq.~\eqref{Eq:off-diag} includes the additional term induced by 
the curl mode: 
\al{
	\ave{\hT_{\bL}\hT_{\bl-\bL}}\rom{CMB,n} 
		= \grad_{\bl} f^{\grad}_{\bl,\bL}
		+ \curl_{\bl} f^{\curl}_{\bl,\bL}
	\,. \label{Eq:off-diag-curl}
}
Even in this case, the estimator for the gradient mode is the same as in Eq.~\eqref{Eq:QE-g}, 
at least, if we consider the first order of the gradient and curl modes. 
This is because the property of the parity symmetry is different for the gradient and curl modes, and 
the inner product $[f^{\grad},f^{\curl}]_{\bl}$, which leads to a bias in the gradient-mode estimator, 
vanishes \cite{Namikawa:2011cs}. 
The quadratic estimators with the polarization anisotropies are also constructed in the same way 
as in the temperature case. 
The optimal quadratic estimator is finally obtained by combining all quadratic combinations of 
temperature and polarization fluctuations with the appropriate weight functions \cite{Hu:2001kj}.

\subsubsection{Practical cases} 

In practical situations, any non-lensing anisotropies arising from 
the masking \cite{vanEngelen:2012va,Namikawa:2012pe,BenoitLevy:2013bc}, 
inhomogeneous map noise \cite{Hanson:2009dr} 
and the beam asymmetry coupled with the scan strategy \cite{Hanson:2010gu}
will also generate the off-diagonal elements in the covariance matrix 
similar to Eq.~\eqref{Eq:off-diag}, 
and the quadratic estimator is biased, i.e., $\ave{\estx_{\bl}}\rom{CMB,n}\not=0$. 
In the case of polarization, there are also several possible sources generating 
mode-couplings such as the temperature to polarization leakage, rotation of polarization basis and so 
on \cite{Shimon:2007au}. 

To see this, let us consider a modulation on temperature anisotropies given by 
\al{
	\hT(\hatn) = (1+\epsilon(\hatn))(\tT(\hatn) + n(\hatn) )
	\,, 
}
where $\epsilon(\hatn)$ may be regarded as the window function, 
inhomogeneity of the optical depth \cite{Gluscevic:2012qv,O'Bryan:2013bea}, 
Doppler boosting \cite{Aghanim:2013suk,Hanson:2009gu}, and so on. 
The off-diagonal covariance of temperature anisotropies in the absence of the curl mode is given by 
at the first order 
\al{
	\ave{\hT_{\bL}\hT_{\bl-\bL}}\rom{CMB,n} 
		= f^{\grad}_{\bl,\bL} \grad_{\bl} 
		+ f^{\epsilon}_{\bl,\bL} \epsilon_{\bl}
	\,. 
}
where $f^{\epsilon}_{\bl,\bL} = \hCTT_{L}+\hCTT_{|\bl-\bL|}$. 
Substituting the above equation to Eq.~\eqref{Eq:QE-g}, we obtain 
\al{
	\ave{\estg_{\bl}}\rom{CMB,n} = \grad_{\bl} + R_{\ell}^{\grad,\epsilon}\epsilon_{\bl} 
	\,, \label{Eq:QE-g-biased} 
}
where the response function, $R_{\ell}^{\grad,\epsilon}$, or in general, $R_{\ell}^{a,b}$, 
is defined as 
\al{
	R_{\ell}^{a,b} = \frac{A^{a,a}_{\ell}}{A^{a,b}_{\ell}} 
	\,; \qquad 
	A_{\ell}^{a,b} = \frac{1}{[g^{a},f^{b}]_{\ell}} 
	\,. \label{Eq:response}
}
The second term of Eq.~\eqref{Eq:QE-g-biased} is called the mean-field bias, and must be corrected. 

One of the methods to correct the mean-field bias is to construct an estimator for $\epsilon_{\bl}$. 
Similar to the lensing estimator, the estimator for $\epsilon_{\bl}$ is constructed using 
the weight function $f_{\bl,\bL}^{\epsilon}$ instead of $f_{\bl,\bL}^{\grad}$. 
The estimator of $\epsilon_{\bl}$ is, however, biased by the presence of lensing as 
\al{
	\ave{\este_{\bl}}\rom{CMB,n} = \epsilon_{\bl} + R_{\ell}^{\epsilon,\grad}\grad_{\bl} 
	\,, \label{Eq:QE-e-biased}
}
where $R_{\ell}^{\epsilon,\grad}$ is defined in Eq.~\eqref{Eq:response}. 
Combining Eqs.~\eqref{Eq:QE-g-biased} and \eqref{Eq:QE-e-biased} to eliminate the term proportional to 
$\epsilon_{\bl}$, we find an unbiased estimator for the gradient mode: 
\al{
	\estg'_{\bl} = 
		\frac{\estg_{\bl} - R_{\ell}^{\grad,\epsilon}\este_{\bl}}
		{1-R_{\ell}^{\grad,\epsilon}R_{\ell}^{\epsilon,\grad}} 
	\,. \label{Eq:QE-g-BHE}
}
Note that the above estimator is derived as the optimal estimator in the case when $\grad_{\bl}$ and 
$\epsilon_{\bl}$ are simultaneously estimated. 

Even if we know the property of $\epsilon_{\bl}$ (e.g., the window function), 
the estimator defined in Eq.~\eqref{Eq:QE-g-BHE} is useful as a cross check of systematics 
propagated from imperfect understanding of underlying CMB anisotropies \cite{Namikawa:2012pe}. 
The similar method can be also applied to reduce the inmohogeneous noise, unresolved point sources, 
polarization angle systematics \cite{Namikawa:2012pe,Ade:2013tyw}, 
as well as for polarization-based reconstruction to reduce bias from the temperature-to-polarization 
leakage, rotation of polarization basis, and so on \cite{Namikawa:2013}.

\subsubsection{Maximum Likelihood Estimator}

Here we comment on the maximum-likelihood estimator of Refs.~\cite{Hirata:2002jy,Hirata:2003ka} 
(see also Ref.~\cite{Hanson:2009kr} for a thorough review). 
Given a set of observed CMB anisotropies, we can formally derive the estimator for the lensing 
fields based on maximizing the likelihood. 
Although the numerical calculation of the maximum-likelihood estimator is difficult compared to 
the quadratic estimator, it is possible to improve the precision of the estimated gradient 
and curl modes. For the gradient mode, 
the expression of the maximum-likelihood estimator is nearly identical to that of the quadratic 
estimator if we only use the temperature anisotropies for the lensing reconstruction 
\cite{Hirata:2002jy}. On the other hand, as shown in Ref.~\cite{Hirata:2003ka}, 
the maximum-likelihood estimator with the B-mode polarization significantly improves the sensitivity 
to the lensing signals, compared to the quadratic estimator. This is because the sensitivity of the 
maximum-likelihood estimator is limited by the intrinsic scatter of the {\it primary} CMB anisotropies 
while the sensitivity of the quadratic estimator is limited by the {\it lensed} CMB anisotropies. 
These situations would be also similar for the curl mode. The quadratic estimator is still useful for 
experiments with the polarization sensitivity of $\gsim 5\mu$K-arcmin which corresponds to the 
amplitude of the B-mode polarization at $\ell\,\lsim 1000$.

\subsection{Estimating CMB lensing power spectrum} 
\label{LensRec:Power}

The angular power spectrum of the gradient and curl modes may be studied through 
the angular power spectrum of the estimators discussed in the previous section. 
The angular power spectrum of the quadratic estimators, however, includes additional contributions 
from, e.g., the four-point correlation of the lensed CMB anisotropies, and 
methods to accurately estimate these bias terms are required. 

To see this, from Eq.~\eqref{Eq:QE-g}, we consider the angular power spectrum 
of the quadratic estimator with temperature which is given by 
\al{
	\ave{|\estx_{\bl}|^2} = (A_{\ell}^x)^2\Int{2}{\bL}{(2\pi)^2}\Int{2}{\bL'}{(2\pi)^2}
		\frac{f^x_{\bl,\bL}}{2\hCTT_{L}\hCTT_{|\bl-\bL|}}
		\frac{f^x_{\bl,\bL'}}{2\hCTT_{L'}\hCTT_{|\bl-\bL'|}}
		\ave{\hT_{\bL}\hT_{\bl-\bL}(\hT_{\bL'}\hT_{\bl-\bL'})^*}
	\,. \label{est-var}
}
This quantity probes the 4-point function of the lensed CMB. Following Eq.~\eqref{Eq:4-point-decomp}, 
we decompose the above quantity into disconnected and connected parts : 
\al{
	\ave{|\hat{x}_{\bl}|^2} 
		= \ave{|\hat{x}_{\bl}|^2}\rom{d} +  \ave{|\hat{x}_{\bl}|^2}\rom{c}
	\,. \label{Eq:pow}
}
The disconnected part, $\ave{\cdots}\rom{d}$, which comes from Eq.~\eqref{Eq:4-point:d}, contains 
the contributions which would be expected if the observed temperature anisotropies  $\hT_{\bL}$ were 
a Gaussian random variable. On the other hand, the connected part, $\ave{\cdots}\rom{c}$, arising from 
Eq.~\eqref{Eq:lens-trispectrum}, has the non-Gaussian contributions which are a distinctive 
signature of lensing. As shown in the following, the connected part nearly corresponds to the power 
spectrum of the gradient/curl mode, and therefore the disconnected part, usually referred to as 
``Gaussian bias'', and the other bias terms must be accurately subtracted to obtain a clean 
measurement of the lensing signals. 

Let us discuss the explicit expression of the disconnected and connected part. 
\bi 
\item {\it Disconnected part} : \\ 
Using Eq.~\eqref{Eq:4-point:d}, the explicit form of the Gaussian bias is written as 
\al{
	\ave{\hT_{\bL}\hT_{\bl-\bL}(\hT_{\bL'}\hT_{\bl-\bL'})^*}\rom{d}
		= (2\pi)^2[\delta_D(\bL+\bL')+\delta_D(\bL+\bL'-\bl)] 
		\hC_L^{\Theta\Theta}\hC_{|\bl-\bL|}^{\Theta\Theta}
	\,. \label{connect-4p}
}
Substituting Eq.(\ref{connect-4p}) into Eq.(\ref{est-var}), and using the expression of the 
normalization $A_{\ell}^{x}$, the power spectrum of the estimator induced by the disconnected part 
becomes 
\al{
	N_{\ell}^{x,(0)} \equiv \ave{|\estx_{\bl}|^2}\rom{d}  
		= \left\{\Int{2}{\bL}{(2\pi)^2}
		\frac{(f^x_{\bl,\bL})^2}{2\hCTT_{L}\hCTT_{|\bl-\bL|}}\right\}^{-1} 
		= A_{\ell}^{x}
	\,. \label{Eq:powD}
}

\item {\it Connected part} : \\
Substituting Eq.~\eqref{Eq:lens-trispectrum} into Eq.~\eqref{est-var}, 
the connected part of the quadratic estimator is, on the other hand, given by 
\cite{Kesden:2003cc}
\al{
	\ave{|\hat{x}_{\bl}|^2}\rom{c} = C_{\ell}^{xx} + N_{\ell}^{x,(1)} 
		+ \mC{O}[(C_{\ell}^{xx})^2]
	\,. \label{Eq:powC}
}
Here $C_{\ell}^{xx}$ is the gradient-mode power spectrum which we wish to estimate, 
while $N_{\ell}^{x,(1)}$ is a nuisance term coming from the ``secondary'' lensing contractions of 
the trispectrum \cite{Hu:2001fa} which is usually called the N1 bias. 

\ei 

\vs{1}

Combining Eqs.~\eqref{Eq:powD} and \eqref{Eq:powC} with Eq.~\eqref{Eq:pow}, we obtain 
\al{
	\ave{|\hat{x}_{\bl}|^2} 
		= C_{\ell}^{xx} + N_{\ell}^{x,(0)} + N^{x,(1)}_{\ell} + \mC{O}[(C_{\ell}^{xx})^2]
	\,. \label{Eq:RecLens:bias}
}
The above equation means that the lensing power spectrum $C_{\ell}^{xx}$ is measured 
by computing the power spectrum of the lensing estimator and subtracting the accurate estimation of 
bias terms such as $N_{\ell}^{x,(0)}$ and $N^{x,(1)}_{\ell}$. 

The Gaussian bias is usually larger than the gradient/curl-mode power spectrum for 
reconstructions with noisy map. 
A method to improve sensitivity to $C_{\ell}^{xx}$ is to use an observed map filtered by 
a realization-dependent power spectrum, instead of its ensemble-averaged quantities 
\cite{Dvorkin:2009ah} in Eq.~\eqref{Eq:powD}. 
In addition, the realization-dependent estimate has an advantage to reduce 
the off-diagonal covariance, 
$\ave{\widehat{C}^{\grad\grad}_{\ell}\widehat{C}^{\grad\grad}_{\ell'}}$ \cite{Hanson:2010rp}. 
For practical situations in which the covariance of observed map has non-negligible off-diagonal 
components, the following estimator is useful as a realization-dependent approach 
\cite{Namikawa:2012pe} 
\al{
	\widehat{N}_{\ell}^{x,(0)} 
		= \left( A_{\ell}^{x} \right)^2\frac{1}{2}\Int{2}{\bL}{(2\pi)^2}\Int{2}{\bL'}{(2\pi)^2}
			f^x_{\bl,\bL} f^x_{\bl,\bL'}
		\left( 2 \ol{\bR{C}}_{ \bL, \bl-\bL' } \bT_{\bl-\bL} \bT_{\bL'}^{*} 
		- \ol{\bR{C}}_{ \bL, \bl-\bL' } \ol{\bR{C}}_{\bl-\bL, \bL'} \right) 
	\,. \label{Eq:BH}
}
Here $\bT_{\bl}\equiv\sum_{\bl'}\bR{C}^{-1}_{\bl,\bl'}\hT_{\bl'}$ is the inverse-variance filtered 
multipoles and $\ol{\bR{C}}_{\bl,\bl'}$ is the covariance of $\bT$. The above estimator is naturally 
derived based on the optimal estimator of trispectrum \cite{Regan:2010cn} applied to lensing 
\cite{Namikawa:2012pe,Ade:2013tyw}, and is easily extended to include polarization 
\cite{Namikawa:2013}. 
Eq.~\eqref{Eq:BH} has an additional advantage for accurate estimation of $C_{\ell}^{xx}$; 
if the covariance is biased as $\ol{\bR{C}}_{\bl,\bl'}\to \ol{\bR{C}}_{\bl,\bl'}+\Sigma_{\bl,\bl'}$, 
contributions of $\Sigma_{\bl,\bl'}$ in Eq.~\eqref{Eq:BH} is at second order, 
while the usual method has the first-order contributions of $\Sigma_{\bl,\bl'}$. 
Another way to mitigate uncertainties in $N^{x,(0)}_{\ell}$ is that, 
since a large fraction of noise in the lensing reconstruction comes from the CMB fluctuations 
themselves, we can construct a Gaussian-bias free estimator by dividing 
the CMB multipoles into disjoint regions in Fourier space, with a cost of signal-to-noise 
\cite{Hu:2001fa,Sherwin:2010ge}. 
For polarization-based reconstructions, the Gaussian bias is more simply mitigated by combining, e.g., 
$EE$ and $EB$ estimator since the four-point correlation $\ave{EEEB}\rom{d}$ vanishes. 

Other bias terms such as the N1 bias $N^{x,(1)}_{\ell}$ should be also corrected. 
Even in the absence of the curl mode, $N^{\curl,(1)}_{\ell}$ is generated by the presence 
of the gradient mode \cite{BenoitLevy:2013bc,vanEngelen:2012va}. 
Furthermore, Ref.~\cite{Hanson:2010rp} pointed out that the term including the second order of 
$C_{\ell}^{\grad\grad}$ in Eq.~\eqref{Eq:RecLens:bias} also leads to non-negligible bias. 
This type of bias can be mitigated by replacing the {\it unlensed} power spectrum 
in the weight function with the {\it lensed} power spectrum \cite{Lewis:2011fk,Anderes:2013jw}. 
The diagonal approximation of the normalization $A_{\ell}$ would also lead to a bias in estimating 
the power spectrum in the presence of, e.g., the window function. 
The bias due to this diagonal approximation in the presence of the masking and survey boundary 
is not so significant for the temperature-based reconstruction \cite{Namikawa:2012pe}, 
but would be significant on large scales for polarization-based reconstruction. 
For known sources such as the window effect, 
we would estimate the normalization bias by Monte Carlo simulations, 
but cross check with other methods would be desirable as a test of assumptions in simulations, 
e.g., underlying CMB anisotropies. 

Since the power spectrum of the quadratic estimator probes the four-point correlation of observed 
anisotropies, other possible sources of the four-point correlation may lead to significant bias on 
$\widehat{C}^{xx}_{\ell'}$. 
One of the significant trispectrum sources is the point sources 
\cite{Ade:2013tyw}, and Ref.~\cite{Osborne:2013nna} constructed an estimator for mitigating 
the point-source trispectrum by modeling the statistical properties of the point sources, 
while Ref.~\cite{vanEngelen:2013rla} proposed a simulation-based approach. 
The bias on estimates of the power spectrum due to the presence of primordial non-Gaussianity 
would be also a source of the trispectrum but is negligible even 
even if $f\rom{NL}\sim\mC{O}(10)$ \cite{Lesgourgues:2005}. 

In estimating cosmological parameters with the gradient/curl-mode power spectrum, the angular power 
spectrum of observed CMB maps is usually added to break degeneracies between parameters. One concern 
in this case is the correlation of the angular power spectrum between lensed CMB and deflection 
angles. Assuming a Planck-like experiment with temperature alone, this correlation is negligible 
\cite{Schmittfull:2013}. The covariance of the angular power spectrum of lensing fields is 
investigated in Refs.~\cite{Kesden:2003cc,Hanson:2010rp}, and is almost diagonal for this case.

\section{Recent experimental progress and future prospect} 
\label{Sec:obs}

\subsection{Current status of observations} 

\begin{figure}[t]
\bc
\includegraphics[width=95mm,clip]{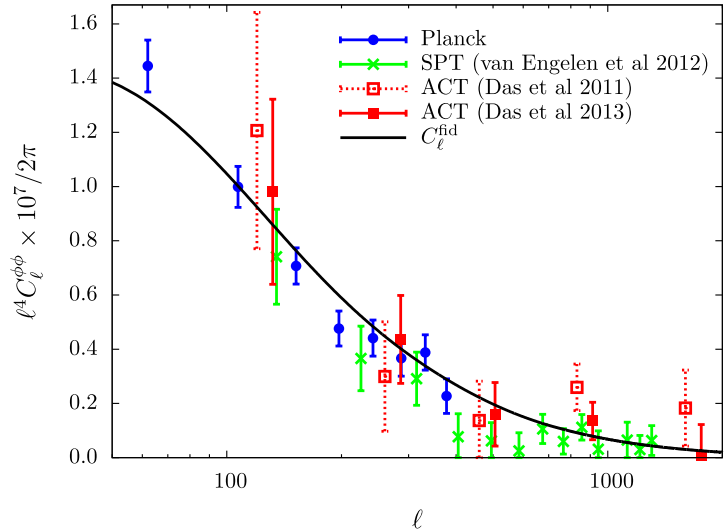}
\caption{
Measurements of the angular power spectrum of the gradient mode obtained 
from Planck \cite{Ade:2013tyw}, SPT \cite{vanEngelen:2012va}, and ACT \cite{Das:2011ak,Das:2013zf}, 
with the temperature-based lensing reconstruction. 
The solid line shows the theoretical power spectrum expected from the best-fit cosmological 
parameters to the Planck temperature data. 
}
\label{fig:ObsResults_Clgg}
\ec
\end{figure}

\begin{figure}[t]
\bc
\includegraphics[width=95mm,clip]{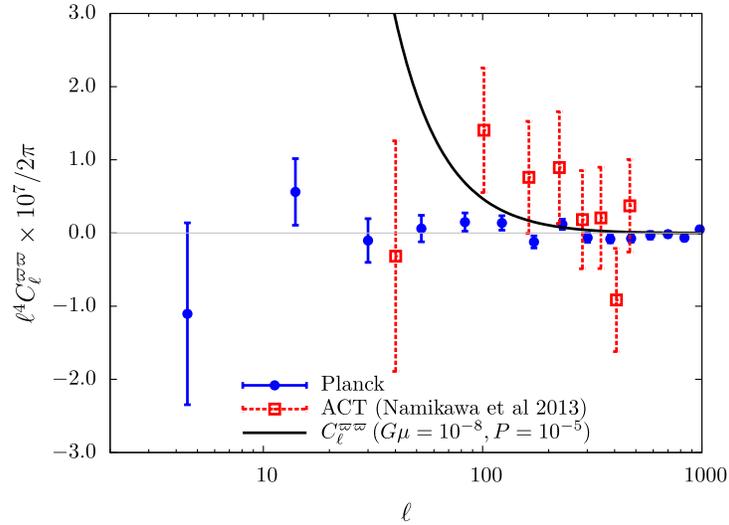}
\caption{
Same as Fig.~\ref{fig:ObsResults_Clgg} but for the curl mode obtained 
from Planck \cite{Ade:2013tyw} and ACT temperature map \cite{Namikawa:2013wda}. 
The solid line shows the theoretical power spectrum for a specific model of cosmic-string network 
with $G\mu=10^{-8}$ and $P=10^{-5}$. 
Note that the curl mode from SPT temperature data is also analyzed in Ref.~\cite{vanEngelen:2012va}. 
}
\label{fig:ObsResults_Clcc}
\ec
\end{figure}


\begin{table}
\bc
\caption{
Current observational status of the measurement of the CMB lensing power spectrum and 
cross power spectrum with other data. 
Each column shows the gradient-mode power spectrum ($\grad\times\grad$), 
the cross-correlation with the galaxy/quasar catalog ($\grad\times G$), 
the cosmic-infrared background map ($\grad\times$CIB) 
or other probes such as the cosmic shear ($\grad\times\gamma$), 
the integrated Sachs-Wolfe effect ($\grad\times$ISW) 
and thermal Sunyaev-Zel'dovich effect ($\grad\times$tSZ). 
Note that part of the results is obtained by combining additional data e.g., WMAP. 
We also note that the lensing effect has been measured from 
the power spectrum of CMB anisotropies with several experiments (see text for details). 
} 
\label{Table:Obs}
\begin{threeparttable}
{\normalsize
\begin{tabular}{c||c||c|c|c} 
	\multicolumn{5}{c}{Temperature} 
	\\ \hline 
	& $\grad\times\grad$ & $\grad\times$G & $\grad\times$CIB & $\grad\times$ other probes 
	\\ \hline 
		ACT
	& 
		$4\,\sigma$ {\small\cite{Das:2011ak}} 
	& 
		$3.8\,\sigma$ {\small\cite{Sherwin:2012mr}} 
	& 
		---
	& 
		$3.2\,\sigma$ {\small\cite{Hand:2013xua}} ($\times\gamma$)
	\\ 
	& 
		$4.6\,\sigma$ {\small\cite{Das:2013zf}} 
	& 
	& 
	& 
	\\ \hline 
		Planck 
	& 
		$26\,\sigma$ {\small\cite{Ade:2013tyw}} 
	& 
		$7\,\sigma$ - $20\,\sigma$ {\small\cite{Ade:2013tyw}} 
	& 
		$42\,\sigma$ {\small\cite{Ade:2013aro}} \tnote{a}
	& 
		$2.5\,\sigma$ {\small\cite{Ade:2013dsi}} ($\times$ISW)
	\\ 
	& 
	& 
	& 
	& 
		$6.2\,\sigma$ {\small\cite{Hill:2013dxa}} ($\times$tSZ)
	\\ \hline 
		SPT 
	& 
		$6.3\,\sigma$ {\small\cite{vanEngelen:2012va}}
	& 
		$4.2$ - $5.3\,\sigma$ {\small\cite{Bleem:2012gm}} 
	& 
		$8.8\,\sigma$ {\small\cite{Holder:2013hqu}} 
	& 
		--- 
	\\ 
	& 
	& 
		$\sim 7\,\sigma$ {\small\cite{Geach:2013zwa}} 
	& 
	& 
		--- 
	\\ \hline 
		WMAP
	& 
		--- \tnote{b}
	& 
		$\sim 3\,\sigma$ {\small\cite{Smith07,Hirata:2008cb,Feng:2012uf}} \tnote{c}
	& 
		---
	& 
		--- \tnote{d} 
	\\ \hline 
	\multicolumn{5}{c}{} 
	\\ \multicolumn{5}{c}{Polarization} 
	\\ \hline 
	& $\grad\times\grad$ & $\grad\times$G & $\grad\times$CIB & $\grad\times$ other probes 
	\\ \hline 
		PolarBear 
	& 
		$\sim 2\,\sigma$ {\small\cite{PB1:2013a}}\tnote{e}
	& 
		--- 
	& 
		$4.0\,\sigma$ {\small\cite{PB1:2013b}} 
	& 
		--- 
	\\ \hline 
		SPTpol 
	& 
		$\sim 3\,\sigma$ {\small\cite{Hanson:2013daa}}\tnote{f}
	& 
		---
	& 
		$7.7\,\sigma$ {\small\cite{Hanson:2013daa}} 
	& 
		--- 
	\\ \hline 
\end{tabular}
}
\begin{tablenotes}\footnotesize 
\item[a] Statistical significance at 545 GHz. 
\item[b] Ref.~\cite{Feng:2011jx} showed that the significance is $\lsim 1$ - $2\,\sigma$. 
\item[c] The measurement of the cross-correlation was first attempted by 
Ref.~\cite{Hirata:2004rp}, but the signals are not detected. 
\item[d] Statistical significance of cross-correlation with the sum of SZ and ISW is at 
$\lsim 1$ - $2\,\sigma$ \cite{Calabrese:2009bu}. 
\item[e] The statistical significance for the rejection of the null hypothesis is at $4.6\,\sigma$. 
\item[f] Ref.~\cite{Hanson:2013daa} constrained the lensing amplitude as a 
consistency test. 
\end{tablenotes}
\end{threeparttable}
\ec
\end{table}

Observations of the lensing effect on CMB are rapidly improving (see Table \ref{Table:Obs}). 
Combining the Arcminute Cosmology Bolometer Array Receiver (ACBAR) with the 
Wilkinson Microwave Anisotropy Probe (WMAP) data, 
Ref.~\cite{Reichardt:2008ay} reported a weak evidence 
of the lensing effect on the temperature power spectrum by constraining a parameter $q$ 
which characterizes the lensing effect as 
$C_{\ell}^{\rm lens}=C_{\ell}^{\rm no-lens}+q(C_{\ell}^{\rm lens}-C_{\ell}^{\rm no-lens})$. 
On the other hand, Ref.~\cite{Calabrese:2009tt} showed a constraint on the lensing amplitude $A$ 
by replacing $C_{\ell}^{\grad\grad}\to AC_{\ell}^{\grad\grad}$ in computing the lensed 
temperature power spectrum, and found $A=3.0^{+0.9}_{-0.9}$ while Ref.~\cite{Reichardt:2008ay} 
showed $A=1.60^{+0.55}_{-0.26}$. 
The lensing effect on the temperature power spectrum has been also explored by 
several high-resolution experiments such as 
the Atacama Cosmology Telescope (ACT) \cite{Das:2010ga} and 
South Pole Telescope (SPT) \cite{Keisler:2011aw,Story:2012wx}. 
The recent Planck result \cite{Ade:2013zuv} 
showed clear evidence for the lensing effect on the temperature 
power spectrum at $\gsim 10\,\sigma$ statistical significance 
\footnote{
Note that $A>1$ is favored at $\sim 2\,\sigma$ \cite{Ade:2013zuv}.
}.
The polarization signals have been also used to show evidence for the lensing effect on 
the CMB anisotropies. 
The recent SPTpol results showed the detection of B-mode polarization signals generated 
from the lensing effect by cross-correlating a map of the cosmic-infrared background obtained 
from the Herschel \cite{Hanson:2013daa}. 
Using the polarization data obtained from the PolarBear experiment, 
the B-mode maps were also used to measure the B-mode angular power spectrum \cite{Ade:2014afa}. 

As shown in Fig.~\ref{fig:ObsResults_Clgg}, 
the power spectrum of the gradient mode obtained through the lensed CMB trispectrum 
has been also explored by several CMB experiments. 
The power spectrum has been measured at $\sim 4\,\sigma-6\,\sigma$ significance 
based on the ACT \cite{Das:2011ak,Das:2013zf} or SPT \cite{vanEngelen:2012va} temperature maps.
At the time of writing this paper, the most precise measurement of the power spectrum 
is given by the Planck with greater than $25\,\sigma$ detection \cite{Ade:2013tyw}. 
CMB polarization maps from the SPTpol \cite{Hanson:2013daa} and PolarBear \cite{PB1:2013a} are also 
utilized to measure the gradient-mode power spectrum. 
As shown in Fig.~\ref{fig:ObsResults_Clcc}, 
the curl-mode power spectrum has been measured with ACT \cite{Namikawa:2013wda}, 
SPT \cite{vanEngelen:2012va} and Planck temperature maps \cite{Ade:2013tyw}, and is consistent 
with zero. 

There are also several efforts to measure the cross correlation between the CMB lensing and 
other observables. 
Cross-correlation with matter density fluctuations is detected at $2$-$3\,\sigma$ significance 
with the data set of WMAP and observations of the large-scale structure such as 
Sloan Digital Sky Survey (SDSS) and NRAO VLA Sky Survey (NVSS) \cite{Smith07,Hirata:2008cb}. 
The first detection of the cross-correlation was made earlier than 
the measurements of the gradient-mode power spectrum. 
The galaxy/quasar-CMB lensing cross-correlation has been also measured by 
Refs.~\cite{Sherwin:2012mr,Bleem:2012gm,Ade:2013tyw}. 
Cross correlations with map of the cosmic-infrared background has been also measured 
and utilized to estimate the bias factor of dusty sources \cite{Holder:2013hqu} and 
the star formation rate \cite{Ade:2013aro}. 
This correlation is more significant than the cross-correlation with 
the galaxy/quasar number density since the cosmic-infrared background is sensitive to the density 
fluctuations of dark matter mostly around $z\sim 2$, corresponding to the peak of 
the CMB lensing kernel \cite{Song:2002sg}. 
Ref.~\cite{Hand:2013xua} reported a measurement of the cross correlation with the cosmic shear 
using data from the ACT and Canada-France-Hawaii Telescope (CFHT) Stripe 82 Survey (CS82). 
Measurements of cross-correlations with other CMB secondaries such as 
the integrated Sachs-Wolfe effect and thermal Sunyaev-Zel'dovich effect are reported 
in Refs.~\cite{Ade:2013dsi} and \cite{Hill:2013dxa}, respectively. 

The lensing signals are now one of the standard probes in cosmology, and have been already 
used for several cosmological issues. 
The inclusion of the gradient-mode power spectrum breaks degeneracies 
of parameters involved in the angular-diameter distance 
to the last scattering \cite{Hu:2001fb}, e.g., 
the dark energy density $\Omega_{\Lambda}$ and curvature parameter $\Omega\rom{K}$, 
whose degeneracies are difficult to break only with the primary CMB anisotropies alone 
\cite{Zaldarriaga:1997ch,Bond:1997wr,Efstathiou:1998xx} (see also about numerical effect 
which breaks degeneracies \cite{Howlett:2012mh}). 
As shown in Refs.~\cite{Sherwin:2011gv,vanEngelen:2012va,Ade:2013tyw}, 
combining the lensing signals with the primary CMB anisotropies provides the evidence for 
dark energy with CMB data alone, 
and the constraints on the dark energy density without any astrophysical data is now 
$\Omega_{\Lambda}=0.67^{+0.027}_{-0.023}\,(1\sigma)$ \cite{Ade:2013tyw}. 
There have been several studies which used the cross correlation to some specific issues. 
Using the cross correlation between the lensing and galaxy survey, constraints on the primordial 
non-Gaussianity parameter through a measurement of the galaxy bias are obtained as 
$f\rom{NL}=12\pm 21\,(1\sigma)$ \cite{Giannantonio:2013kqa}. 
Ref.~\cite{Ferraro:2014msa} used the Planck lensing map to 
constrain the bias of Wide-Field Infrared Survey Explorer (WISE) for the purpose of 
estimating the ISW effect. 
As discussed in Sec.~2, the curl mode of lensing signals has also fruitful information on 
the non-scalar perturbations. 
Fig.~\ref{fig:ObsResults_Clcc} shows the measured curl-mode power spectrum 
compared with that produced by the cosmic-string network. 
The measured curl-mode power spectrum is used for excluding parameter region of 
cosmic strings \cite{Namikawa:2013wda}
which is not ruled out by the current data of the temperature power spectrum 
\cite{Ade:2013xla,Yamauchi:2011cu}.

\subsection{Future prospects} 

CMB polarization data on arcminute scale will soon become the best way to obtain the CMB lensing 
power spectrum and cross-correlations, and these precise signals play an important role in cosmology 
in near future 
(see e.g., Ref.~\cite{Wu:2014hta} and Refs therein). 
This will be achieved by ongoing ground-based experiments such as 
ACTPol \footnote{\url{http://www.princeton.edu/act/}}, 
PolarBear \footnote{\url{http://bolo.berkeley.edu/polarbear/}}, 
and 
SPTpol \footnote{\url{http://pole.uchicago.edu/}}, 
and upcoming/next generation experiments, e.g., 
Polar \footnote{\url{http://polar-array.stanford.edu/}}, 
CMBPol \footnote{\url{http://cmbpol.uchicago.edu/}}, 
COrE\footnote{\url{http://www.core-mission.org/}} 
and 
PRISM \footnote{\url{http://www.prism-mission.org/}}. 

Based on the above planned experiments, let us discuss the future prospect in CMB lensing studies. 
For the neutrinos, assuming upcoming/next-generation experiments 
and combining the gradient-mode power spectrum with the primary CMB power spectrum, 
$1\,\sigma$ constraints on the total mass of neutrinos would be 
$35$-$60\,{\rm meV}$ (e.g., \cite{Kaplinghat:2003bh,Lesgourgues:2005yv,dP09,Namikawa:2010re}). 
On the other hand, the $1\,\sigma$ constraint on the effective number of neutrinos will be $\sim 0.1$ 
\cite{Lesgourgues:2005yv}. 

The cross-correlation studies with the CMB lensing would also become important in the future. 
Inclusion of the cross correlations with other observables will further improve the constraints 
on the total mass of the neutrinos (e.g., \cite{Abazajian:2011dt,Abazajian:2013oma} and Refs therein). 
For example, if we combine the Stage-II class experiments with other ongoing projects such as 
the Subaru Hyper Suprime-Cam \footnote{\url{http://www.naoj.org/Projects/HSC/index.html}}, 
the constraints on neutrino mass would be $40\,{\rm meV}$ \cite{Namikawa:2010re}. 
In the future, with the Stage-IV class experiments and other upcoming spectroscopic survey, 
the constraint on the mass of neutrinos and effective number of neutrinos would be 
$\sim 16\,{\rm meV}$ and $0.02$, respectively \cite{Abazajian:2013oma}. 
This implies that the lower bound on the neutrino mass $\sum m_{\nu}\sim 60\,{\rm mV}$ 
obtained from neutrino oscillation experiments would be detected 
at $4\,\sigma$ confidence level with future experiments. 

For upcoming and future experiments, the auto and cross power spectrum between CMB lensing 
and other observables would have sensitivity to probe the dark-energy equation-of-state parameters 
(e.g., \cite{Hu:2001fb,Das:2008am,Namikawa:2010re}), 
a specific model of dark energy (e.g., \cite{Das:2008am,dP09})/modified gravity 
(e.g., \cite{Calabrese:2009tt}), 
the primordial non-Gaussianity through measurement of galaxy bias 
(e.g., \cite{Takeuchi:2011ej}), 
and the cosmic-string network (e.g., \cite{Yamauchi:2012bc}). 
In addition to probe the above advanced issues, 
the cross correlation with other probes would help to control systematics 
such as multiplicative bias and intrinsic alignment in the cosmic shear analysis 
(e.g., \cite{Vallinotto:2011ge,Das:2013aia,Hall:2014nja,Troxel:2014kza}). 

In the future, as mentioned in Sec.~\ref{Rev:CMBlensing}, the delensing may be required to 
obtain the primary B-mode signals at the recombination bump ($\ell\sim 10-100$). 
The B-mode signal at these scales would be important for ground-based experiments in which 
the large-scale modes are difficult to obtain. 
The delensing at the recombination bump is also important for future low-resolution 
space missions such as LiteBIRD \footnote{\url{http://cmbpol.kek.jp/litebird/index.html}} and 
PIXIE \cite{Kogut:2011xw} in order to enhance the total signal-to-noise of the primary B-mode 
as well as the sensitivity to the tensor spectral index which allows to explore 
the details of inflationary physics. 
The joint analysis for, e.g., LiteBIRD and ground-based CMB experiments would reveal 
the primordial B-mode signals from the largest scale to the recombination bump, providing us 
with much information on the primordial gravitational waves. 
The above estimates and prospects are however discussed in simple and idealistic situations, 
and studies aiming at addressing practical issues are highly required as data become precise.


\section*{Acknowledgments}
TN thanks Duncan Hanson, Ryo Nagata, Atsushi Taruya and Daisuke Yamauchi for helpful 
comments on this review, greatly appreciates the Planck team for kindly providing us with 
the curl-mode power spectrum, acknowledges the use of {\tt CAMB} \cite{Lewis:1999bs}, 
and would like to thank the anonymous referees for improving the text. 
TN is supported in part by JSPS Grant-in-Aid for Research Activity Start-up (No. 80708511). 
Numerical computations were carried out on SR16000 at YITP in Kyoto University and Cray XT4 at Center 
for Computational Astrophysics, CfCA, of National Astronomical Observatory of Japan.

\appendix

\bibliographystyle{ptephy}
\bibliography{cite}

\begin{thebibliography}{100}

\bibitem{Abazajian:2011dt}
K.~N. Abazajian et~al., Astropart. Phys., {\bf 35}, 177--184 (2011).

\bibitem{Abazajian:2013vfg}
K.~N. Abazajian et~al., {{arXiv:1309.5381}}.

\bibitem{Abazajian:2013oma}
K.~N. Abazajian et~al., {{arXiv:1309.5383}}.

\bibitem{Anderes:2013jw}
E.~Anderes, Phys. Rev. D (2013).

\bibitem{BenoitLevy:2013bc}
A.~Benoit-Levy et~al., Astron. Astrophys., {\bf 555}, 10 (2013).

\bibitem{BenoitLevy:2012va}
A.~Benoit-Levy, K.~M. Smith, and W.~Hu, Phys. Rev., {\bf D86}, 123008 (2012).

\bibitem{Bernardeau:1996aa}
F.~Bernardeau, Astron. Astrophys., {\bf 324}, 15--26 (1997).

\bibitem{Blanchard:1987AA}
A.~Blanchard and J.~Schneider, Astron. Astrophys., {\bf 184}, 1--6 (oct 1987).

\bibitem{Bleem:2012gm}
L.~E. Bleem et~al., Astrophys. J., {\bf 753}, L9 (2012).

\bibitem{1980:Bond}
J.~R. Bond, G.~Efstathiou, and J.~Silk, Phys. Rev. Lett., {\bf 45}, 1980--1984
  (1980).

\bibitem{Bond:1997wr}
J.~R. Bond, G.~Efstathiou, and M.~Tegmark, Mon. Not. Roy. Astron. Soc., {\bf
  291}, L33--L41 (1997).

\bibitem{Calabrese:2009tt}
E.~Calabrese et~al., Phys. Rev. D, {\bf 80}, 103516 (2009).

\bibitem{Calabrese:2009bu}
E.~Calabrese et~al., Phys. Rev. D, {\bf 81}, 043529 (2010).

\bibitem{Carvalho:2010rz}
C.~S. Carvalho and K.~Moodley, Phys. Rev. D, {\bf 81}, 123010 (2010).

\bibitem{Challinor:2005jy}
A.~Challinor and A.~Lewis, Phys. Rev. D, {\bf 71}, 103010 (2005).

\bibitem{Ade:2013dsi}
Planck Collaboration, {{arXiv:1303.5079}}.

\bibitem{Ade:2013zuv}
Planck Collaboration, {{arXiv:1303.5076}}.

\bibitem{Ade:2013tyw}
Planck Collaboration, {{arXiv:1303.5077}}.

\bibitem{Ade:2013aro}
Planck Collaboration, {{arXiv:1303.5078}}.

\bibitem{Ade:2013xla}
Planck Collaboration, {{arXiv:1303.5085}}.

\bibitem{Aghanim:2013suk}
Planck Collaboration, {{arXiv:1303.5087}}.

\bibitem{PB1:2013b}
PolarBear Collaboration, {{arXiv:1312.6645}}.

\bibitem{PB1:2013a}
PolarBear Collaboration, {{arXiv:1312.6646}}.

\bibitem{Ade:2014afa}
PolarBear Collaboration, {{arXiv:1403.2369}}.

\bibitem{Cooray:2002mj}
A.~Cooray and W.~Hu, Astrophys. J., {\bf 574}, 19 (2002).

\bibitem{Cooray:2005hm}
A.~Cooray, M.~Kamionkowski, and R.~R. Caldwell, Phys. Rev. D, {\bf 71}, 123527
  (2005).

\bibitem{Cooray:2008xz}
A.~Cooray, D.~Sarkar, and P.~Serra, Phys. Rev. D, {\bf 77}, 123006 (2008).

\bibitem{Dai:2012bc}
L.~Dai, M.~Kamionkowski, and D.~Jeong, Phys. Rev. D, {\bf 86}, 125013 (2012).

\bibitem{Das:2013aia}
S.~Das, J.~Errard, and D.~Spergel, {{arXiv:1311.2338}}.

\bibitem{Das:2010ga}
S.~Das et~al., Astrophys. J., {\bf 729}, 62 (2011).

\bibitem{Das:2011ak}
S.~Das et~al., Phys. Rev. Lett., {\bf 107}, 021301 (2011).

\bibitem{Das:2013zf}
S.~Das et~al., {{arXiv:1301.1037}}.

\bibitem{Das:2008am}
S.~Das and D.~N. Spergel, Phys. Rev. D, {\bf 79}, 043509 (2009).

\bibitem{dP09}
R.~de~Putter, O.~Zahn, and E.~V. Linder, Phys. Rev. D, {\bf 79}, 065033 (2009).

\bibitem{Dodelson:2003bv}
S.~Dodelson, E.~Rozo, and A.~Stebbins, Phys. Rev. Lett., {\bf 91}, 021301
  (2003).

\bibitem{Dvorkin:2009ah}
C.~Dvorkin, W.~Hu, and K.~M. Smith, Phys. Rev. D, {\bf 79}, 107302 (2009).

\bibitem{Efstathiou:1998xx}
G.~Efstathiou and J.~R. Bond, Mon. Not. Roy. Astron. Soc., {\bf 304}, 75--97
  (1999).

\bibitem{Feng:2012uf}
C.~Feng et~al., Phys. Rev. D, {\bf 86}, 063519 (2012).

\bibitem{Feng:2011jx}
C.~Feng et~al., Phys. Rev. D, {\bf 85}, 043513 (2012).

\bibitem{Ferraro:2014msa}
S.~Ferraro, B.~D. Sherwin, and D.~N. Spergel, {{arXiv:1401.1193}}.

\bibitem{Fukushige:1994}
T.~Fukushige, J.~Makino, and T.~Ebisuzaki, Astrophys. J., {\bf 436}, L107--L110
  (dec 1994).

\bibitem{Geach:2013zwa}
J.E. Geach et~al., Astronomical Journal, {\bf 776}, L41 (2013).

\bibitem{Giannantonio:2013kqa}
T.~Giannantonio and W.~J. Percival, {{arXiv:1312.5154}}.

\bibitem{Gluscevic:2012qv}
V.~Gluscevic, M.~Kamionkowski, and D.~Hanson, {{arXiv:1210.5507}}.

\bibitem{Goldberg:1999xm}
D.~M. Goldberg and D.~N. Spergel, Phys. Rev. D, {\bf 59}, 103002 (1999).

\bibitem{Guzik:2000ju}
J.~Guzik, U.~Seljak, and M.~Zaldarriaga, Phys. Rev. D, {\bf 62}, 043517 (2000).

\bibitem{Hall:2014nja}
A.~Hall and A.~Taylor, {{arXiv:1401.6018}}.

\bibitem{Hand:2013xua}
N.~Hand et~al., {{arXiv:1311.6200}}.

\bibitem{Hanson:2009kr}
D.~Hanson, A.~Challinor, and A.~Lewis, Gen. Rel. Grav., {\bf 42}, 2197--2218
  (2010).

\bibitem{Hanson:2009kg}
D.~Hanson et~al., Phys. Rev. D, {\bf 80}, 083004 (2009).

\bibitem{Hanson:2010rp}
D.~Hanson et~al., Phys. Rev. D, {\bf 83}, 043005 (2011).

\bibitem{Hanson:2013daa}
D.~Hanson et~al., Phys. Rev. Lett., {\bf 111}, 141301 (2013).

\bibitem{Hanson:2009gu}
D.~Hanson and A.~Lewis, Phys. Rev. D, {\bf 80}, 063004 (2009).

\bibitem{Hanson:2010gu}
D.~Hanson, A.~Lewis, and A.~Challinor, Phys. Rev. D, {\bf 81}, 103003 (2010).

\bibitem{Hanson:2009dr}
D.~Hanson, G.~Rocha, and K.~Gorski, Mon. Not. Roy. Astron. Soc., {\bf 400},
  2169--2173 (2009).

\bibitem{Hill:2013dxa}
J.~C. Hill and D.~N. Spergel, {{arXiv:1312.4525}}.

\bibitem{Hirata:2004rp}
C.~M. Hirata et~al., Phys. Rev. D, {\bf 70}, 103501 (2004).

\bibitem{Hirata:2008cb}
C.~M. Hirata et~al., Phys. Rev. D, {\bf 78}, 043520 (2008).

\bibitem{Hirata:2002jy}
C.~M. Hirata and U.~Seljak, Phys. Rev. D, {\bf 67}, 043001 (2003).

\bibitem{Hirata:2003ka}
C.~M. Hirata and U.~Seljak, Phys. Rev. D, {\bf 68}, 083002 (2003).

\bibitem{Holder:2013hqu}
G.~P. Holder et~al., Astrophys. J., {\bf 771}, L16 (2013).

\bibitem{Howlett:2012mh}
C.~Howlett et~al., JCAP, {\bf 1204}, 027 (2012).

\bibitem{Hu:2000ee}
W.~Hu, Phys. Rev. D, {\bf 62}, 043007 (2000).

\bibitem{Hu:2001fa}
W.~Hu, Phys. Rev. D, {\bf 64}, 083005 (2001).

\bibitem{Hu:2001}
W.~Hu, Astrophys. J., {\bf 557}, L79--L83 (2001).

\bibitem{Hu:2001fb}
W.~Hu, Phys. Rev. D, {\bf 65}, 023003 (2002).

\bibitem{Hu:2001yq}
W.~Hu and A.~Cooray, Phys. Rev. D, {\bf 63}, 023504 (2001).

\bibitem{Hu:2001kj}
W.~Hu and T.~Okamoto, Astrophys. J., {\bf 574}, 566--574 (2002).

\bibitem{Hu:1997hp}
W.~Hu and M.~J. White, Phys. Rev. D, {\bf 56}, 596--615 (1997).

\bibitem{Kaplinghat:2003bh}
M.~Kaplinghat, L.~Knox, and Y.-S. Song, Phys. Rev. Lett., {\bf 91}, 241301
  (2003).

\bibitem{Keisler:2011aw}
R.~Keisler et~al., Astrophys. J., {\bf 743}, 28 (2011).

\bibitem{Kesden:2003cc}
M.~H. Kesden, A.~Cooray, and M.~Kamionkowski, Phys. Rev. D, {\bf 67}, 123507
  (2003).

\bibitem{Knox:2002pe}
L.~Knox and Y.-S. Song, Phys. Rev. Lett., {\bf 89}, 011303 (2002).

\bibitem{Kodama:1984}
H.~Kodama and M.~Sasaki, Progress of Theoretical Physics Supplement, {\bf 78},
  1 (1984).

\bibitem{Kogut:2011xw}
A.~Kogut et~al., JCAP, {\bf 1107}, 025 (2011).

\bibitem{Lesgourgues:2005}
J.~Lesgourgues et~al., Phys. Rev. D, {\bf 71}, 103514 (2005).

\bibitem{Lesgourgues:2005yv}
J.~Lesgourgues et~al., Phys. Rev. D, {\bf 73}, 045021 (2006).

\bibitem{Lewis:2006fu}
A.~Lewis and A.~Challinor, Phys. Rep., {\bf 429}, 1--65 (2006).

\bibitem{Lewis:2011fk}
A.~Lewis, A.~Challinor, and D.~Hanson, JCAP, {\bf 1103}, 018 (2011).

\bibitem{Lewis:1999bs}
A.~Lewis, A.~Challinor, and A.~Lasenby, Astrophys. J., {\bf 538}, 473--476
  (2000).

\bibitem{Li:2006pu}
C.~Li, T.~L. Smith, and A.~Cooray, Phys. Rev. D, {\bf 75}, 083501 (2007).

\bibitem{Namikawa:2012pe}
T.~Namikawa, D.~Hanson, and R.~Takahashi, Mon. Not. Roy. Astron. Soc., {\bf
  431}, 609--620 (2013).

\bibitem{Namikawa:2010re}
T.~Namikawa, S.~Saito, and A.~Taruya, JCAP, {\bf 1012}, 027 (2010).

\bibitem{Namikawa:2013}
T.~Namikawa and R.~Takahashi, Mon. Not. Roy. Astron. Soc. (2013).

\bibitem{Namikawa:2011cs}
T.~Namikawa, D.~Yamauchi, and A.~Taruya, JCAP, {\bf 1201}, 007 (2012).

\bibitem{Namikawa:2013wda}
T.~Namikawa, D.~Yamauchi, and A.~Taruya, Phys. Rev. D (2013).

\bibitem{O'Bryan:2013bea}
J.~O'Bryan et~al., {{arXiv:1306.1232}}.

\bibitem{Okamoto:2002ik}
T.~Okamoto and W.~Hu, Phys. Rev. D, {\bf 66}, 063008 (2002).

\bibitem{Okamoto:2003zw}
T.~Okamoto and W.~Hu, Phys. Rev. D, {\bf 67}, 083002 (2003).

\bibitem{Osborne:2013nna}
S.~J. Osborne, D.~Hanson, and O.~Dore, {{arXiv:1310.7547}}.

\bibitem{Regan:2010cn}
D.~M. Regan, E.~P.~S. Shellard, and J.~R. Fergusson, Phys. Rev. D, {\bf 82},
  023520 (2010).

\bibitem{Reichardt:2008ay}
C.~L. Reichardt et~al., Astrophys. J., {\bf 694}, 1200--1219 (2009).

\bibitem{Sarkar:2008ii}
D.~Sarkar et~al., Phys. Rev. D, {\bf 77}, 103515 (2008).

\bibitem{Sasaki:1989}
M.~Sasaki, Mon. Not. Roy. Astron. Soc., {\bf 240}, 415--420 (1989).

\bibitem{Schmalzing:2000}
J.~Schmalzing, M.~Takada, and T.~Futamase, Astrophys. J., {\bf 544}, L83--L86
  (2000).

\bibitem{Schmidt:2012nw}
F.~Schmidt and D.~Jeong, Phys. Rev. D, {\bf 86}, 083513 (2012).

\bibitem{Schmittfull:2013}
M.~M. Schmittfull et~al., Phys. Rev. D, {\bf 88}, 063012 (2013).

\bibitem{Seljak:1995ve}
U.~Seljak, Astrophys. J., {\bf 463}, 1 (1996).

\bibitem{Seljak:2003pn}
U.~Seljak and C.~M. Hirata, Phys. Rev. D, {\bf 69}, 043005 (2004).

\bibitem{Seljak:1998nu}
U.~Seljak and M.~Zaldarriaga, Phys. Rev. D, {\bf 60}, 043504 (1999).

\bibitem{Seljak:1998aq}
U.~Seljak and M.~Zaldarriaga, Phys. Rev. Lett., {\bf 82}, 2636--2639 (1999).

\bibitem{Serra:2008wc}
P.~Serra and A.~Cooray, Phys. Rev. D, {\bf 77}, 107305 (2008).

\bibitem{Sherwin:2010ge}
B.~D. Sherwin and S.~Das, {{arXiv:1011.4510}}.

\bibitem{Sherwin:2011gv}
B.~D. Sherwin et~al., Phys. Rev. Lett., {\bf 107}, 021302 (2011).

\bibitem{Sherwin:2012mr}
B.~D. Sherwin et~al., Phys. Rev. D, {\bf 86}, 083006 (2012).

\bibitem{Shimon:2007au}
M.~Shimon et~al., Phys. Rev. D, {\bf 77}, 083003 (2008).

\bibitem{Smith:2011we}
K.~M. Smith, ASP Conf. Ser., {\bf 432}, 147 (2009).

\bibitem{Smith:2008an}
K.~M. Smith et~al., AIP Conf. Proc., {\bf 1141}, 121 (2009).

\bibitem{Smith:2010gu}
K.~M. Smith et~al., JCAP, {\bf 1206}, 014 (2012).

\bibitem{Smith:2006nk}
K.~M. Smith, W.~Hu, and M.~Kaplinghat, Phys. Rev. D, {\bf 74}, 123002 (2006).

\bibitem{Smith07}
K.~M. Smith, O.~Zahn, and O.~Dore, Phys. Rev. D, {\bf 76}, 043510 (2007).

\bibitem{Smith:2002dz}
R.~E. Smith et~al., Mon. Not. Roy. Astron. Soc., {\bf 341}, 1311 (2003).

\bibitem{Smith:2005ue}
S.~Smith, A.~Challinor, and G.~Rocha, Phys. Rev. D, {\bf 73}, 023517 (2006).

\bibitem{Song:2002sg}
Y.-S. Song et~al., Astrophys.J., {\bf 590}, 664--672 (2003).

\bibitem{Stebbins:1996wx}
A.~Stebbins, {{arXiv:astro-ph/9609149}}.

\bibitem{Story:2012wx}
K.~T. Story et~al., Astrophys. J., {\bf 779}, 86 (2013).

\bibitem{Takada:2001b}
M.~Takada, Astrophys. J., {\bf 558}, 29--41 (2001).

\bibitem{Takada:2001a}
M.~Takada and T.~Futamase, Astrophys. J., {\bf 546}, 620--634 (2001).

\bibitem{Takada:2000}
M.~Takada, E.~Komatsu, and T.~Futamase, Astrophys. J., {\bf 533}, L83--L87
  (2000).

\bibitem{Takahashi:2012em}
R.~Takahashi et~al., Astrophys. J., {\bf 761}, 152 (2012).

\bibitem{Takeuchi:2011ej}
Y.~Takeuchi, K.~Ichiki, and T.~Matsubara, Phys. Rev. D, {\bf 85}, 043518
  (2012).

\bibitem{Teng:2011xc}
W.-H. Teng, C.-L. Kuo, and J.-H.~P. Wu, {{arXiv:1102.5729}}.

\bibitem{Tomita:1989}
K.~Tomita and K.~Watanabe, Progress of Theoretical Physics, {\bf 82}, 563--580
  (sep 1989).

\bibitem{Troxel:2014kza}
M.~A. Troxel and M.~Ishak, {{arXiv:1401.7051}}.

\bibitem{Vallinotto:2011ge}
A.~Vallinotto, Astrophys. J., {\bf 759}, 32 (2012).

\bibitem{vanEngelen:2012va}
A.~van Engelen et~al., Astrophys. J., {\bf 756}, 142 (2012).

\bibitem{vanEngelen:2013rla}
A.~van Engelen et~al., {{arXiv:1310.7023}}.

\bibitem{Verde:2005ff}
L.~Verde, H.~Peiris, and R.~Jimenez, JCAP, {\bf 0601}, 019 (2006).

\bibitem{Wilkinson:2013kia}
R.~J. Wilkinson, J.~Lesgourgues, and C.~Boehm, {{arXiv:1309.7588}}.

\bibitem{Wu:2014hta}
W.L.K. Wu et~al., {{arXiv:1402.4108}}.

\bibitem{Yamauchi:2011cu}
D.~Yamauchi et~al., Phys. Rev. D, {\bf 85}, 103515 (2012).

\bibitem{Yamauchi:2012bc}
D.~Yamauchi, T.~Namikawa, and A.~Taruya, JCAP, {\bf 1210}, 030 (2012).

\bibitem{Yamauchi:2013fra}
D.~Yamauchi, T.~Namikawa, and A.~Taruya, JCAP, {\bf 1308}, 051 (2013).

\bibitem{Yamauchi:2013pna}
D.~Yamauchi, Y.~Sendouda, and K.~Takahashi, {{arXiv:1309.5528}}.

\bibitem{Zaldarriaga:2000ud}
M.~Zaldarriaga, Phys. Rev. D, {\bf D62}, 063510 (2000).

\bibitem{Zaldarriaga:1998ar}
M.~Zaldarriaga and U.~Seljak, Phys. Rev. D, {\bf 58}, 023003 (1998).

\bibitem{Zaldarriaga:1998te}
M.~Zaldarriaga and U.~Seljak, Phys. Rev. D, {\bf 59}, 123507 (1999).

\bibitem{Zaldarriaga:1997ch}
M.~Zaldarriaga, D.~N. Spergel, and U.~Seljak, Astrophys. J., {\bf 488}, 1--13
  (1997).

\end{thebibliography}

\end{document}